# The economics of malnutrition:
# Dietary transition and food system transformation

William A. Masters[1*], Amelia B. Finaret[2] and Steven A. Block[3]

## Abstract
Rapid increases in food supplies have reduced global hunger, while rising burdens of diet-related disease have made poor diet quality the leading cause of death and disability around the world. Today's "double burden" of undernourishment in utero and early childhood then undesired weight gain and obesity later in life is accompanied by a third less visible burden of micronutrient imbalances. The triple burden of undernutrition, obesity, and unbalanced micronutrients that underlies many diet-related diseases such as diabetes, hypertension and other cardiometabolic disorders often coexist in the same person, household and community. All kinds of deprivation are closely linked to food insecurity and poverty, but income growth does not always improve diet quality in part because consumers cannot directly or immediately observe the health consequences of their food options, especially for newly introduced or reformulated items. Even after direct experience and epidemiological evidence reveals relative risks of dietary patterns and nutritional exposures, many consumers may not consume a healthy diet because food choice is driven by other factors. This chapter reviews the evidence on dietary transition and food system transformation during economic development, drawing implications for how research and practice in agricultural economics can improve nutritional outcomes.

## Keywords
Food policy, nutrition transition, diet quality, diet-related disease

## Affiliations
1. Friedman School of Nutrition and Department of Economics, Tufts University
2. Department of Global Health, Allegheny College
3. The Fletcher School and Department of Economics, Tufts University
* Corresponding author:  william.masters@tufts.edu

## Acknowledgments
We are grateful to Chris Barrett and David Just for the invitation and guidance in writing this chapter, to John Hoddinott and other participants in the AAEA Handbook of Agricultural Economics workshop for early feedback, as well as comments on a later draft from the editors and two anonymous reviewers. Special thanks to Shelly Sundberg and Winnie Bell for detailed suggestions, to Laurian Unnevehr and Wally Falcon for their guidance, and to the countless other researchers, students and survey respondents from whom we have learned so much and will continue to learn in the years ahead.

**The economics of malnutrition:**
**Dietary transition and food system transformation**

## Contents

## Figures

**The economics of malnutrition:**
**Dietary transition and food system transformation**

## Introduction

More than two centuries after Malthus and fifty years after the Green Revolution, changes in global agriculture and food systems remain central to human and economic development. Large increases in global food supplies have alleviated much but not all the world's hunger, while unbalanced intake of macro- and micro-nutrients, interacting with poor sanitation and lack of health care, have made malnutrition the world's leading avoidable cause of death and disability (Murray et al. 2020).

This chapter describes the multiple burdens of malnutrition that affect human health, and reviews their links to dietary transition, agriculture, and food system transformation. The many forms of malnutrition include dietary imbalances that lead to growth faltering in early childhood, followed by unwanted weight gain and lifelong higher risk of diabetes, hypertension or other cardiometabolic disease as well as some cancers and vulnerability to infectious disease. A "triple burden" of early-life undernourishment, later weight gain, and ongoing diet-related diseases associated with poor diet quality often coexist in the same person, household and community, with broad patterns of transition in the prevalence of each condition over time and space (Barrett and Bevis 2015; Pinstrup-Andersen 2007).

Some forms of malnutrition are closely linked to poverty and food insecurity, leading to health disparities that are associated with deprivation, disempowerment, and social exclusion. Other aspects of malnutrition are more common at middle- or higher income levels and may be driven by factors other than household income (Brown, Ravallion and Van De Walle 2017). Our focus in this chapter is malnutrition in terms of diet quality for health, derived from the nutrient composition and other bioactive attributes of foods consumed. These dietary factors often cannot be detected by consumers before, during or after eating, and their impacts on health have only recently become known due to large-scale studies over long periods of time (Mozaffarian, Rosenberg and Uauy 2018). Related concerns such as food insecurity (defined as not always having enough money or other resources to acquire the quantity or quality of foods desired), food safety (defined as avoidance of unwanted contaminants), and the many cultural or culinary dimensions of food that enter consumer demand all influence the nutritional quality of the diet. These effects on health are then mediated by a variety of additional factors such as physical activity, the gut microbiome and enteric function, parasitic diseases and other aspects of human physiology and metabolism over the life course (Finaret and Masters 2019).

Examples of undernutrition linked to insufficient nutrient intake interacting with disease include stunting, defined as height-for-age less than 2 SD below the median of a healthy population, and anemia defined as hemoglobin count below 0.11 g/ml of blood, adjusted for altitude. Using the most recent nationally representative Demographic and Health Surveys (DHS) data for each country in the StatCompiler of ICF International (2012), prevalence of stunting among children under 5 years of age ranged from 6.9 percent in the Dominican Republic in 2013 to 55.9 percent in Burundi in 2016, while the prevalence of anemia among children under age 5 ranged from 15.6 percent in Albania in 2015 to 86.3 percent in Yemen in 2013 (ICF International 2012).

Examples of malnutrition caused by overconsumption of some nutrients, including obesity and diet-related disease, often worsen with economic development especially when new foods with unhealthy attributes are introduced to the food supply. Overweight prevalence in children under 5 years, defined by the DHS as being 2 SD above the median weight-for-height in a healthy population at each age, ranged in prevalence from 0.9 percent in Senegal in 2017 to 16.4 percent in Albania in 2017. Among adult women, the prevalence of overweight or obesity ranged from 7.6 percent in Ethiopia (3.5 percent for men) in 2016 to 62.2 percent in South Africa in 2016. Prevalence rates of overweight or obesity among men are lower, at 3.5 percent in Ethiopia in 2016 and 27.4 percent in South Africa in 2016 (ICF International 2012).

A central theme of this chapter is that the early-life origins of many chronic diseases creates opportunities for highly cost-effective preventive interventions and policies, especially in maternal and child nutrition (Barker 1990, Daniels 2016, Currie and Rossin-Slater 2015). This chapter also reviews the evidence on dietary transition and food system transformation during economic development, drawing implications for how research, policy, and practice in agricultural economics can improve nutritional outcomes. Reviewing the evidence on how economic development relates to malnutrition can inform many aspects of agricultural economics research and policy, guiding change towards improved nutritional outcomes. We update and build upon a nutrition economics foundation built by many, especially several previous Handbook chapters with a nutrition or food intake focus such as Huffman and Orazem (2007), Barrett (2002), and Antle (2001).

Section 1 of the chapter provides a summary of how nutritional status is measured for research and policy purposes, and a summary of economic explanations for food intake and dietary choice. Section 2 reviews the epidemiological data and trends on the nutrition transition in dietary intake and associated health outcomes in relation to economic transformation in agriculture and the economy as a whole. Section 3 reviews the data and trends on dietary change and the food system transformation over time and across countries and regions. Section 4 presents evidence on policies to address various forms of malnutrition, and Section 5 concludes with a discussion of future research directions and causal inference in the economics of nutrition.

## 1. Measurement of dietary intake and health outcomes

The role of dietary intake in health outcomes has been debated since antiquity and is often misunderstood. Effects may play out over months and years, with so many confounding factors that false beliefs can arise and persist for centuries. Even the phrase "let food be thy medicine", often considered part of the foundational Hippocratic Corpus from around 400 BC, appears to be a later invention that contradicts Hippocrates' other writing, which aims to distinguish medical intervention from water, food and other naturally occurring factors affecting health (Cardenas 2013). Popular opinion and medical authorities often involve mistaken beliefs about even the most basic functions of food, from ancient ideas about humors (Wilkins 2020) to recent reversals in understanding of how dietary fats and cholesterol affect body composition and metabolism (Mozaffarian, Rosenberg and Uauy 2018).

Despite the prevalence of false beliefs about nutrition, there is abundant historical evidence from many diverse populations that higher-income people consistently consumed diets that helped them grow taller and live longer (Fogel 2004, Floud et al. 2011, Deaton 2013). The nutritional value of food includes some attributes that are directly visible to consumers and could be supplied by competing producers in unregulated markets. Other nutrition-related attributes are experience goods whose value is known to the consumer after purchase or credence goods whose value remains unknown even after consumption, so that sales depend on consumer's beliefs and trust in a brand or a third-party certifier (Akerlof 1970). Food markets have a very long history of collective action to signal otherwise unobservable quality dating back to Britain's Assize of Bread introduced by King John in 1202, Bavaria's beer purity laws of 1516, and many other voluntary or mandatory grades and standards (Swinnen 2016). Over time the evolution of food culture and regulations could deliver increasingly safe and healthy food as incomes grew, but it is now clear that higher income no longer consistently leads to improved nutrition as shown for example by puzzling declines in the attained height of apparently well-nourished populations (Pannett 2021).

Progress towards improved diet quality could stall for many reasons, including the recent introduction of food attributes that are appealing but ultimately harmful in ways that cannot be discerned through trial and error or previous research. Scientific discovery of diet-disease relationships is constrained by the high cost and relatively short durations of randomized trials, the limited availability and selective focus of research funding, and the difficulty of causal inference from observational epidemiological studies (Hörnell et al. 2017). The impact of each item added or removed from a person's diet includes how it affects other foods and behaviors, leading to a focus on broad dietary patterns rather than individual foods in both observational studies (Schulze et al. 2018) and randomized trials (e.g. Hall et al. 2019). Regardless of these challenges, carefully designed and implemented observational studies as well as randomized trials can provide meaningful evidence on how dietary patterns affect health (Satija et al. 2018).

New evidence from nutrition research can help economists model food demand and consumption. Some aspects of food quality can be detected by consumers or credibly signaled by individual producers, while others depend on quality assurance from third parties such as voluntary associations or public authorities. The publication of new research findings can itself shift demand, often in ways that complicate the relationship. For example, when credible evidence for health benefits of whole grains was added to the *Dietary Guidelines for Americans* in 2005, demand for whole-grain products rose but only among higher-income consumers (Mancino and Kuchler 2012). As shown by Oster (2020), nutritional recommendations lead to increased demand primarily among health-seeking consumers who pursue other inputs to health as well, adding spurious correlation to any underlying causal effect of the food itself. Even when health effects of dietary choices are fully known and observable, many factors other than health could drive food choice. Over time, innovations in technology and regulation can help people meet those other objectives of food consumption related to taste and satiation, convenience and cultural identity in healthier ways, while also pursuing goals for food production such as improved environmental quality and animal welfare, reduced carbon and methane emissions, or the well-being of food system workers. The resulting change in dietary intake leads to systematic patterns of nutrition transition as described below.

## 1.1 Nutrition transition in dietary intake and health outcomes

The term "nutrition transition" was coined by food economist Barry Popkin and his collaborators in the 1990s (Popkin 1994; Drewnowski and Popkin 1997), describing changes in dietary intake and health outcomes in terms of five stages. In the first stage, which accounts for most of human evolution, people were hunter-gatherers with diverse diets leading to tall but lean bodies and a high risk of infectious disease. The second stage of Popkin's nutrition transition occurs with domestication of starchy staples as the population grows and dietary deficiencies led to shorter statures due to poor maternal and child health and low intake of protein, healthy fats and micronutrients. In the third stage, improvements in living standards lead to greater heights and improved health for more affluent consumers, but there is continued "hidden hunger" especially for women and children with greater physiological need for certain nutrients, and growing health disparities due to the intergenerational transmission of malnutrition from gestation and birth through childhood and adulthood. Many hidden deficiencies at this stage are invisible without clinical examination and testing, and are remedied through health services and nutritional intervention. In a fourth stage of transition, food manufacturers introduce new formulations and marketing efforts that lead to unwanted weight gain and a growing burden of non-communicable cardiometabolic diseases such as diabetes and hypertension as well as renal disorders and some cancers. The fifth and final stage of the nutrition transition happens when diet-related risks are discovered and remedied through changes in the food environment and consumer behavior.

The five stages of Popkin's nutrition transition may occur sequentially but may also coexist in some populations to varying degrees. A simplified framework describes these changes in terms of a "double burden" in anthropometric status relating to short stature as a marker of undernourishment, and obesity as a marker of overconsumption (Wells et al. 2019). For this chapter, the transition is extended to multiple burdens that include diet-related diseases at every level of height and weight, and more generally to "malnutrition in all its forms" (Hawkes et al. 2020). Overall, nutrition can be seen as a Goldilocks problem, moving from undernutrition to overconsumption of each dietary component before reaching an overall healthy diet. Optimal ranges for each aspect of the diet are initially unknown, and are discovered only after some overshooting reveals the boundaries of both deficiency and excess for each nutrient or other bioactive compound in the diet (Meenakshi 2016).

Reducing the burdens of undernutrition across each stage of the transition would provide substantial economic benefits, with specific nutrition investments having benefit-cost ratios that range from 3.6 to 48.0 in low- and middle-income countries (Alderman et al. 2017; Hoddinott et al. 2013). Key steps towards just-right nutrition include income growth or safety nets so that households can acquire enough foods with health attributes that are visible and well-known, but also changes in food technology, institutions, and policy for the many attributes of food whose health effects are unknown to the consumer. The economics of nutrition hinge on the interaction between effective demand by individuals and households, incentives and technology that drive food supply, and policy interventions, public investments, and regulations that can improve the outcome of interaction between demand and supply (Finaret and Masters 2019, 2020).

*Costs of malnutrition: disease and loss of productivity*
Unlike acute illness caused by specific pathogens in the absence of food safety, diseases caused by malnutrition often develop slowly through long-term exposure to interacting risk factors. For example, macronutrient imbalances often cause obesity and diabetes, and excess sodium often causes hypertension, both of which involve gradual changes in metabolism and cardiovascular physiology whose onset is slow and may be difficult to detect. These and other diet-related harms may be treatable but are rarely fully reversed, so without preventive measures they either go untreated leading to illness, disability and death, or they impose a high burden of treatment for those receiving care.

While malnutrition is most often discussed in terms of noncommunicable disease, some forms of malnutrition are known to also increase risk and severity of infectious disease. The most globally prevalent examples of disease worsened by malnutrition are respiratory infections such as COVID-19 and tuberculosis. For example, about 25% of the world's population have latent tuberculosis in their bodies at any one time (Cohen et al. 2019), but whether infection progresses into active tuberculosis and then the severity of that disease depends greatly on nutritional status as a driver of immune function (Baum, Tamargo and Wanke 2021). Similar relationships may exist for COVID-19 as suggested by close correlations between obesity and its severity, and specific nutrients ranging from vitamin A to zinc deficiency have been associated with the severity of many other infectious diseases (Humphries, Scott and Vermund 2021).

The many forms of malnutrition are clearly linked to a wide range of health outcomes. In an early example of modern research on the economics of nutrition, Strauss and Thomas (1998) wrote that "…health is fundamentally multidimensional, and it is important to differentiate among these dimensions." (p. 768). Building on that observation, in the following section we provide an outline of how each kind of malnutrition is measured.

**1.2 Measuring nutrition: Diet quality and health outcomes**
For clarity of exposition, in this chapter we describe nutritional status in terms of macronutrients, micronutrients, and then a third category for all other bioactive compounds such as antioxidants, probiotics, antinutrients, toxins that can affect health positively or negatively. The term "nutrients" usually refers to either macronutrients (protein, fats and carbohydrates) which supply dietary energy and also perform other functions, or micronutrients which are either minerals or vitamins deemed "essential" for health in the sense that they must be obtained from the diet, and that they are responsible for specific biological functions in the body. Example health outcomes from insufficient or excessive intake in all three dimensions are provided in Table 1 below.

**Table 1. Three dimensions of dietary intake and example effects on health outcomes**

| Type of compound | Example effects of diet quality on human health | |
| | Excess intake | Insufficient intake |
| --- | --- | --- |
| *Macronutrients (protein, fats and carbohydrates)* | Unwanted weight gain, overweight and obesity; insulin resistance and diabetes | Low birthweight and stunted linear growth; underweight and wasting; insufficient weight gain in pregnancy and poor gestational health |
| *Micronutrients (vitamins and minerals)* | Excess sodium contributes to hypertension; Vitamin A is teratogenic in high doses | Vitamin A deficiency causes blindness and mortality from infections; iodine deficiency causes goiter and neurological impairment |
| *Other compounds in food* | Anti-nutrients inhibit absorption; contaminants and mycotoxins cause disease | Bioactive compounds like phytochemicals have anti-cancer properties; whole grains and fermented foods can promote gut health |

For all three dimensions of dietary intake shown in the first column of Table 1, the timing, extent and duration of deficiency or excess can play a major role in malnutrition. Dietary requirements vary over the life course, and bodies may have some nutrient reserves or other mechanisms to maintain homeostasis over a range of intake levels. Malnutrition is therefore often measured with respect to discrete lower and upper bounds that define the healthy range, partly to reflect the realities of physiology, and partly because clinical thresholds are needed for diagnosis and treatment of individuals and populations. Thresholds often differ by age, sex, and reproductive status such as pregnancy and lactation, and may also vary with body size and physical activity. Similar diet quality criteria are typically used universally across all populations because genetic patterns vary more within groups than between groups. Differences in diet-disease outcomes between populations are typically explained by shared environmental conditions, even as new research identifies genetic differences among individuals in their susceptibility to diet-related disease (Loos and Yeo 2021).

Every dimension of dietary intake can vary independently to some degree, with high or low levels of each individual nutrient and other attributes based on the combination of foods being consumed. For most of human history, the nutritional composition of each food item was fixed by the natural process of crop or livestock production, but agricultural transformation and food manufacturing has made for much greater variation in nutrient content and other attributes of the diet. Some of that variation is unintentional, such as change in micronutrient composition when breeding for higher yield or loss of bioactive compounds in milling for refined grains, while other steps are intentional such as enrichment and fortification to add micronutrients, or formulating foods to increase their palatability and quantities consumed. Dietary transition towards packaged and processed foods can be associated with improved health (Debela et al. 2020) or poorer health (Pries et al. 2019, Popkin et al. 2020) depending on how each attribute of the diet is affected by processing and marketing.

The existence of distinct individual nutrients in food was among the first important discoveries of Western science in the 18th and 19th centuries. For macronutrients, the idea that food contains energy was confirmed by Antoine Lavoisier around 1770, followed by the identification of protein, fats, and carbohydrates by Justus Liebig around 1840. For micronutrients, the first known link was between citrus and scurvy that led to isolation of vitamin C, followed by various others such as the link between polished rice and pellagra that led to discovery of vitamin B3 (niacin), and ultimately what is now a total of 13 distinct essential vitamins and about 20 essential minerals whose intake in foods is required for human growth and development. Vitamins are organic molecules that can be destroyed by exposure to heat and light, while minerals are inorganic molecules that can be found on the periodic table of elements. Countless other compounds that also affect health are continually being discovered even today, thanks to improved measurement and analytical methods (Webster-Gandy 2020). Bioactive compounds that are not essential nutrients include carotenoids, flavonoids, phytosterols and polyphenols. These compounds may be found in high concentrations most often in fruits and vegetables, fermented foods or tea and coffee, but starchy staples can also be important sources of healthy attributes such as anthocyanins in maize (Poole et al. 2021).

Following the discovery of each food attribute, clinical trials and epidemiological evidence have attempted to identify lower and upper bounds of intake to limit disease risks. This begins with total quantity of each macro- and micronutrient, and often also addresses the distinct type of protein in term of its component amino acids, the type of fat in terms of saturated, monounsaturated, polyunsaturated, or trans fats, whether carbohydrates are simple or complex and accompanied by fiber (soluble or insoluble). Nutrition discoveries have sometimes revealed that ancient culinary practices are health promoting, such as how combining rice and beans or chapati and dal provides complementary amino acid profiles needed for a complete protein, and how consuming a diversity of food groups improves micronutrient adequacy. In those cases, revealed preferences lead to increasingly healthy diets as incomes rise. At the same time, some deeply rooted human preferences such as choosing refined starches instead of whole grains now have harmful consequences, and the entirely new food components such as trans fats can be attractive to consumers but unknowingly harm their future health. In those situations, increased income and innovation leads to more diet-related disease. As the unhealthy effects of particular kinds of food processing and marketing are discovered, opportunities arise for regulation to mandate disclosure and limit harmful activities while accelerating adoption of healthier practices.

Nutritional status reflects the balance between intake and requirements for many different dietary components, mediated by a wide range of non-dietary factors. Modern measurement techniques for nutritional assessment of individuals are detailed in the textbook by Gibson (1990, 2005, 2021), and summarized in a convenient mnemonic as the ABCDs of nutrition measured in terms of Anthropometry, Biochemistry, Clinical signs and symptoms, and Dietary intake (Dwyer et al. 1993). For modern research on the economics of malnutrition we extend this classification to a larger ABCDEFG set of variables listed below. The list starts with anthropometry used to assess the classic "double burden" of poor childhood growth and then excess weight gain, followed by biomarkers, clinical signs and symptoms, and dietary recall data about intake, which can be extended to metrics of environmental factors like exposure to toxins and pollutants, then food safety, and finally governance of the food system. At the end of the chapter, we will return to this list to describe advancements in data collection and methods for these factors that will enhance economic analyses in the future.

A. **Anthropometric measurement** of a person's body size begins with weight and height and transformed ratios of the two such as the body mass index (BMI). For children, stunting (short stature) is defined as height-for-age more than 2 SD below the median of a healthy population at each age and sex, while wasting (thinness) is defined as a weight-for-height more than 2 SD below the median of a healthy population at each age and sex. For adults, the primary metric is BMI (weight divided by height squared), for which the conventional thresholds define underweight or overweight and obesity as a BMI below 18.5 or above 25 and 30 kg/m$^2$ respectively. Other anthropometric measures that are used to study nutritional status and population health include a child's head circumference to measure skeletal growth, skinfold thickness, and circumference at the waist, hip or upper arm to measure fat deposition at those locations. Body composition regarding bone density, muscle mass and adiposity can also be measured with imaging techniques, displacement of water or electrical impedance. Accelerometers to measure physical activity are closely related to body size, and data from accelerometers in wristbands or mobile phones are increasingly used in conjunction with anthropometry to estimate total energy requirements and other aspects of nutrition and health.

B. **Biochemical indicators** derived from blood, urine, stool or tissue samples help provide a more complete picture of an individual's nutritional status (Brown et al. 2021). For example, hemoglobin concentrations to measure anemia can now be measured at very low cost in field settings. Other biochemical innovations in nutrition assessment include photoelectric measurement of blood oxygen levels, and genetic analysis of stool samples to measure composition of the gut microbiome. For some micronutrients, like calcium, concentrations in the blood are tightly controlled by homeostasis, so other approaches involving urine samples might be needed. Typically, biomarkers measure single dietary factors, but efforts are underway to identify integrative nutritional biomarkers which focus on nutritionally regulated biomarkers of health. The most commonly used biomarkers for nutrition care in high-income countries are cholesterol and triglycerides to indicate cardiovascular health, fasting blood glucose to indicate problems with glucose metabolism, and blood urea nitrogen and creatinine to indicate kidney function.

C. **Clinical signs and symptoms** such as discolored nails, neuropathy, fatigue, or impaired night vision can provide early indications of specific micronutrient deficiencies. Like anthropometry and biochemical measures, these indicators are often integrated into primary care medical practice for early detection before deficiencies lead to more severe illness and disability. Clinical investigation often combines anthropometry and biomarkers, such as bone densitometry plus blood and urine tests to assess osteomalacia and osteoporosis. Clinical research also uses specialized facilities such as metabolic chambers that account for all inflow and outflow of energy and nutrients, allowing for experimental variation and monitoring of factors that cannot otherwise be measured.

D. **Dietary assessment** is practiced by dietitians, nutritional epidemiologists, and trained survey enumerators, using a variety of techniques to overcome the difficulty of remembering and reporting what was eaten with sufficient accuracy to estimate nutrient intake. Contemporaneous food diaries or automated recording is usually infeasible so methods focus on dietary recall, asking qualitative (yes/no) and sometimes quantitative (weight or volume) questions about broad food groups or specific items eaten over the previous day and night. Standard practices call for two 24-hr recalls on different days, followed by a set of data transformations to convert responses into estimated usual intakes. For longer-term dietary pattern analysis, food frequency questionnaires are designed to elicit intake patterns over a week, month, or year. In some cases, inferences about food consumption are made from aggregate data such as national food balance sheets for agricultural commodities, or purchase and sale data for differentiated products. From the tradition of farm surveys in low-income countries, total food consumption is often measured for the whole household at once over the previous 30 days, and some ask about the frequency of consumption over the past 7 days. Increasing interest in diet quality is leading to investment in new survey methods for 24-hour dietary recall which measure the quantity of each food consumed and as well as short diet quality indicators using yes/no questions about specific foods. Added complexity arises for child nutrition, as intake of breastmilk and other foods is especially difficult for caregivers to estimate, particularly given the small quantities, rapid change over time, and volatility in intake across days and weeks as children grow and vary in their calorie needs.

E. **Environmental** and social factors extend the original ABCD framework of nutrition assessment for individuals as defined by Dwyer et al. (1993), but have long been considered important for public health. These include environmental sources of bacteria, viruses and parasites linked to sanitation, airborne toxins and particulates from kitchen smoke or industrial pollution. Other local, regional, national or global variables are often described as social-ecological factors or social determinants of health, with increasing attention to the commercial determinants of health (Allen 2021). Economic aspects of the nutritional environment include food business decisions that shape consumer preferences such as marketing and distribution efforts that place different products more conveniently at hand, or investments in advertising and cultural factors that influence consumers' aspirations, as well as changes in employment and wages, time use, childcare, housing, transportation and other factors that affect a household's food acquisition and meal preparation. Interest in these aspects of food systems has brought efforts to standardize measurement of the degree to which a 'food desert' lacks healthy foods, or a 'food swamp' has too many unhealthy items. Most recently, the prices of diverse foods have been added up using the least-cost items needed to meet each nutritional standard, as a kind of food price index to track the affordability of healthy diets in local markets at each place and time (Masters et al. 2018, Herforth et al. 2020).

F. **Food system** metrics, including farming methods, food safety and food processing, food waste and other variables are not traditionally included in either individual nutritional assessment or public health nutrition, but are of great interest in the economics of malnutrition. We extend the ABCDE mnemonic to food system variables, starting with farming conditions that drives the composition of raw agricultural commodities, which are then transformed by food businesses into a wide variety of food items with varying attributes. Food safety and food waste are particularly important for perishables that are prone to contamination and spoilage, with high levels of biological contaminants such as salmonella, e. coli, or molds that introduce mycotoxins. Traditional food processing techniques often aim to extend shelf life. Examples include milling flour into refined grain to avoid oxidation of the lipids in whole grains, adding sugar and salt to limit microbial growth, or using nitrites to preserve processed meats. Such approaches create important tradeoffs between food safety and a food's nutritional value, while also leading to demand for refrigeration and other ways to preserve and enhance food composition. Once diverse foods are available, food system metrics often focus on agricultural production of nutritional food groups, specifically the quantities of starchy staples (both cereals and starchy roots), leguminous grains (beans and pulses) as well as nuts and seeds, fruits and vegetables of various kinds, animal-sourced foods (meat, fish, eggs and dairy), plus vegetable oils and sugar as ingredients in other foods. The relative proportions of these food groups, plus other concerns such as limits on processed meats to prevent cancer, salt to prevent hypertension, added sugar to prevent cardiovascular disease, and use of whole grains instead of refined flour, are all communicated in national dietary guidelines and widely used for assessment of diet quality of individuals and populations. Another aspect of food systems of great importance for nutrition is food insecurity, usually defined as variation in resources needed to obtain food, measured using a set of internationally standardized questions such as whether the respondent skipped meals, went to bed hungry or ate different foods due to lack of resources to acquire their desired foods.

G. **Governance** factors are a final addition to the list, extending our mnemonic to include political and institutional processes that are of particular interest for economists. Collective action and government regulations, or lack thereof, have shaped food markets throughout human history. Contemporary research on the economics of nutrition increasingly focuses on tracking the policy and program interventions needed to inform consumers and address market failures, such as labeling and disclosure of food composition, mandates for healthy fortification like iodine in salt, bans on harmful ingredients like trans fats, or enforcement of production standards like hazard analysis and critical control point (HACCP) systems. Even when regulations are in place compliance can vary widely, as explored by Ebata et al. (2021) and Saha et al. (2021) for edible oils and salt in Bangladesh. The effects of fortification and enrichment policies, such as use of folic acid in enriched grains, depends not only on legislation but also the frequency of visits from regulators (Saha et al. 2021). The development of governance metrics is in its infancy but identifying best practices to align food components with consumer demand and societal health is essential for the final stage of nutrition transition, when food choices reliably reflect consumer willingness and ability to pay for their immediate needs and future well-being.

The ABCDEFG mnemonic categorization of nutritional metrics can be linked back to the three dimensions of dietary intake summarized in Table 1. Accurately measuring each variable, for individuals or households and entire populations, is challenging but essential to research that might explain the variation in outcomes we observe. Beyond understanding what people consume, there is increased interest in also measuring how and why people choose to consume the foods they do (Blake et al. 2021). To promote good nutrition, dietary guidelines and meal planning can inform food choice for everyone, accommodating the diversity in preferences and traditions across and within populations. Diets should have variety and contain sufficient but not excess nutrients for an active and healthy life. This Goldilocks principle is our starting point for a microeconomic overview of nutrition, which has long focused on estimating budget shares and elasticities for agricultural commodities but has shifted more recently to a focus on the nutritional attributes of retail food items.

### 1.3 The microeconomics of food choice: Revealed preferences and consumer welfare
The economics of nutrition begins with food choice, including microeconomic theory that aims to explain and predict behavior in ways that might allow us to infer well-being from observed levels of consumption. In this section, we review the microeconomic theory and development economics literature that addresses whether observed choices reveal individual preferences and social welfare, and what other information could be used to guide food and nutrition policy. Empirical work on these questions remains limited by data quality and availability, but the fundamental questions are as old as economics itself.

A central question for microeconomic theory is whether observing revealed preferences, when consumers repeatedly choose one option instead of another, implies that the chosen option actually delivers a higher level of well-being (Samuelson 1938). As described in this chapter, observed food choices are guided by biochemical, physiological, psychological and social factors that may not yield consistent rankings at all, let alone a ranking that aligns with the consumer's own lifelong well-being. This chapter focuses on well-being in terms of lifelong health, but other dimensions of well-being are also important, including pleasures obtained from an enjoyable meal, the convenience and predictability of foods that allow consumers to do things other than meal preparation, and even the well-being obtained from signaling social identity through food choice (Atkin et al. 2021).

To infer well-being from observed food choices, economists posit the existence of an unobservable but internally consistent utility function by which different outcomes can be ranked. The idea of a "well behaved" utility function under which rankings are unambiguous is a construct developed and used to estimate how entire populations of consumers will respond to changes in income, prices and a variety of other changes. These changes may or may not reveal which food would provide a higher level of well-being, if only because many important components of each food and their various functions cannot be observed and are not accurately signaled by the food itself, or by any labels and other information available to consumers.

The microeconomic theory by which consumer behavior might be used to infer well-being is explained in texts such as Varian (1978), Deaton (1992) and Mas-Colell et al. (1995). Each principle is introduced graphically in two dimensions for example using indifference curves and budget lines, then generalized to many dimensions using calculus, and ultimately generalized to all possible cases using real analysis and other mathematics. Many applications of consumer theory directly address food and nutrition issues, notably Deaton and Muellbauer (1980) and Behrman, Deolalikar, and Wolfe (1988). For readers interested in a thorough textbook approach to nutrition economics specifically, Babu et al. (2016) provides a clear and concise exposition on theory and empirical applications.

The assumptions needed for revealed preferences to reflect consumers' well-being may or may not hold for foods or bundles of foods. Reviewing these aspects of consumer demand reveals the degree to which social welfare be represented using revealed preferences, and when other approaches would be needed for economic analysis of nutrition and health. When applying demand systems to food and nutrition, economists should consider whether each assumption about preferences is applicable to their research context, to choose the most appropriate model specification.

Microeconomic theory concerns the preconditions under which utility functions might be useful to explain and predict behavior, including the axioms from which to derive functions to estimate the parameters of a multidimensional demand system and its two-dimensional indifference curves. The axioms for preferences to be represented in this way include completeness, reflexivity, transitivity, and continuity, as well as monotonicity, non-satiation, and convexity of the utility function. Whether the completeness axiom holds for food choice depends not only on food composition but also on how that is likely to interact with other determinants of the consumer's health and well-being, as different circumstances may lead to preference reversals. In addition, satisfying the completeness axiom depends on the availability of information and the quality of information about the possible bundles.

Reflexivity is usually described as a trivial axiom, but nutrition economists should note that each food can have a very different value at different times. The time-specific nature of food demand is due not only to changing circumstances around each consumption occasion, but also to differences in discount rates between immediate and future consumption. Empirical examples demonstrating the value of foods at different times include Shapiro (2005), who uses data from SNAP recipients in the U.S. and finds a strong present bias affecting food consumption in general, and Ubfal (2016), who finds that discount rates vary by type of food in Uganda, where consumers are the most impatient about sugar, meat, and plantains.

The transitivity property ensures internal consistency of preferences. For foods, preference reversals can occur for many reasons, including shifts in the salience of given food attributes for choice in any given setting. For example, a consumer primed to focus on taste might prefer beef to chicken and chicken to tofu, but when thinking about environmental harms their preference ordering might switch to tofu, chicken, then beef, and when primed to think about animal welfare that same person's ordering might be tofu then beef and ahead of chicken. Any change in the salience of the three nonmarket attributes could lead to preference reversal and self-contradictory choices, violating the transitivity axiom.

Some food preferences involve adjusting quantities as a continuous function of price and income, but food choices often involve discontinuities or indivisibilities such as switching between food categories and dietary patterns. Food preferences may also be non-monotonic, as there is no free disposal of attributes like sodium for which excesses are harmful. Even when food choices might be consistent with other axioms, the actual composition of foods and its impact on health is often unknown to the consumer, limiting the degree to which revealed preference signals consumer well-being. Finally, some aspects of food choice may not involve voluntary choices at all, as they could also reflect a variety of involuntary conditions and disorders (Miljkovic 2020).

A fundamental aspect of preferences is substitution and complementarity among foods. When two items are perfect substitutes, they meet identical needs so any combination of the two would be equally valuable. When they are perfect complements, they would be consumed in fixed proportions. In both cases, the two foods could enter demand systems as a single item made up of either or both components. Substitution and complementarity is of interest because each food meets diverse needs in different ways, leading to high but imperfect substitutability between similar foods such as diverse cereal grains, which may all be imperfect complements to other kinds of food such as leguminous grains and pulses. Careful empirical estimates of own- and cross-price elasticities can elicit whether foods are substitutes or complements in a given market and setting.

Attributes of foods can also be examined for substitution or complementarity, as in Meas et al. (2016) who demonstrated that "organic" and "local" attributes are substitutes in the market for blackberry jam, while the attribute "comes from a small farm" is a substitute for both "organic" and "local." Examining WTP or the degree of substitution and complementarity for various types of attributes is particularly complex. Choice experiments cannot include the same number of varied attribute choices as the real world, raising questions about the accuracy of the choice experiments (Caputo et al. 2017). Caputo et al. (2017) examine how consumers process different types of attributes, such as "cue" attributes that convey information otherwise unobservable, or "independent" attributes which describe physical properties of the food item.

When utility functions are separable, researchers can focus on a subset of things, such as food choices, independently of variation in choices over other goods and services. Separability is particularly convenient for modeling two- or three-stage budgeting, where consumers first decide how much to allocate towards broad categories like food, healthcare, and housing, and then decide how to allocate within the groups such as for meat, eggs, and starchy foods. If items are closely related, for example if housing decisions have implications for the cost of food purchases and meal preparation, separability should not be assumed.

Separation between consumption categories is based on preference independence, whereby the chosen quantity of one good does not depend on quantities consumed of some other good. Most foods are not independent choices in that sense, but imposing separation may be helpful for estimation and modeling. For example, Clements and Si (2016) use preference independence to derive compensated price elasticities from the uncompensated estimates reported by Green et al. (2013) and Cornelsen et al. (2015), using additional data on budget shares and income elasticities. This can be useful for population-level estimates, even if the assumption would not hold for individual diets.

Some preferences might be internally consistent and yet not conform to a utility function at all. A classic example is a ranking where choice depends first on the quantity of one good, and the quantity of others would matter only as a tie breaker. Lexicographic preferences are not continuous and allow no substitution between goods. Meenakshi et al. (2012) find evidence of lexicographic preferences in a sub-sample of their study population in Zambia, where they estimate willingness-to-pay for orange (Vitamin A rich) maize using a choice experiment. Without accounting for the structure of preferences in their modeling, WTP would be overstated because the lexicographic preference was to choose orange maize only, regardless of the price (Meenakshi et al. 2012). Berg and Preston (2017) find evidence of lexicographic preferences among people in New Zealand who are unwilling to purchase non-local foods regardless of the price. While lexicographic preferences may occur within food groups or sources (i.e., across different attributes like color, sourcing or taste), it is unlikely that people have lexicographic preferences for calories versus non-calories, as explored in work on nutrition-based efficiency wages by Powell and Murphy (2015).

Economic models represent selected aspects of human behavior. Like any model, these simplifications are far from a complete picture, leaving much variance unexplained and possibly introducing systematic biases relative to the unobservable underlying concepts we seek to understand. In this section, we review some of the ways that utility functions are used to explain and predict observed food choice, and some of the limits on the degree to which revealed preference corresponds to unobservable well-being. Readers interested in behavioral economics approaches can refer to Caputo and Just (2022) in this Handbook volume which explores behavioral economics approaches and welfare analysis of nutrition policies in detail.

Food choices are influenced by many psychological factors in addition to the physiological and economic aspects of nutrition. An explosion of research in behavioral economics has focused on understanding the internal decision processes of consumers, and how those decisions might be affected by phenomena such as loss aversion, inequality aversion, herd behavior, risk perceptions, and time consistency. In his essay *Behavior and the Concept of Preference*, Amartya Sen argues that many choices "may be taken on the basis of incomplete thinking about the possible courses of action… (or due to) open or hidden persuasion involved in advertisements and propaganda, which frequently mess up not only one's attitude towards the alternatives available but also towards the act of choice itself" (Sen 1973, p. 247). Food choices may be particularly prone to "incomplete thinking", so that choices made on one occasion are subsequently regretted, or framed in ways that change over time.

Regretted effects of a person's choices on their own future self are sometimes known as "internalities", by analogy to externalities that affect other people, and may be especially frequent in food choices due to biological or psychological compulsions from which a person would prefer to be protected. For example, a food's image or aroma may trigger over-eating, so a person might prefer to avoid being exposed to that kind of health risk. Even more often, however, incomplete thinking is due to information failures, because people simply do not know what compounds are in each food, or how those compounds might affect their long-term health. There may be asymmetric information when food producers might have more knowledge than consumers (in which case third-party quality assurance would be needed) but many compounds have health consequences that have only recently been discovered.

Just and Gabrielyan (2016) argue that the role of information in food choice means that behavioral aspects of economics is essential for measuring policy impact, especially when internalities and information failures are heterogeneous across types of consumers. With attention paid to individual behavioral constraints, economists can explore a rich set of policy options beyond transfers, taxes, regulation and public goods provision to many other aspects of choice architecture needed to make each choice consistent with each consumer's own long-term preferences. Overcoming each kind of market failure is challenging for many reasons, including the role of homeostasis or physiological "set points" such that improvements at one place or time may be offset by compensatory behavior elsewhere, and the social nature of preferences such that improvements for one person may revert to group norms.

The many limitations of revealed preference as a guide to nutritional well-being imply that economists should increasingly use other information about food composition and its effects on health, including metrics derived from the ABCDEFG mnemonic used to study nutrition and public health, to complement the demand system estimation described below.

## 1.4 Empirical analysis of food choice and nutrition
The healthiness of food is not always reflected in effective demand for many reasons, but most fundamentally because consumers cannot observe food composition or the long-term health consequences of each item, and even if health consequences are known, consumption may be driven by other factors such as taste, convenience and advertising. Aspects of diet quality that may worsen despite income growth involve added sugar and other simple carbohydrates, sodium and industrial trans fats as well as removal of dietary fiber and other changes that can make processed foods more attractive but contribute to long-term cardiometabolic disease. Harmful effects of some food attributes have only recently been discovered, such as trans fats, and societies differ in their policy response given new scientific evidence, leading to patchy and uneven progress towards improved diet quality.

This section will review the literature on demand systems, with a focus on the nutritional aspects of food demand. Estimating demand systems allows for context-specific evidence on how changes to incomes, prices, and preferences will change demand, which is essential for policy analysis. The parameters to be estimated include income elasticities, own-price and cross-price elasticities (compensated and uncompensated), along with their consequences for budget shares and quantities consumed at each level of income and prices. Shifts in preferences would alter these parameters, due for example to a change in product formulation or marketing that alters taste, convenience and other factors that might affect food choice at each level of income and prices. Those shifts include new information about how food choices contribute to consumers' nutritional status as an input to their health production function, or otherwise affect consumers' subjective well-being, identity and social status.

Statistical estimation of revealed preferences uses demand systems derived from well-behaved utility functions. The estimated equations would correspond directly to social welfare only if each individual choice actually maximized well-being, and did so under a utility function that is sufficiently stable and uniform to estimate a single set of demand system parameters for that population over time. More generally, we can think of demand system parameters as extracting one aspect of the data, namely that part of the variance which predicts food choice on average for a population of interest. This can be very useful for nutrition policy. For example, using a complete demand system and a utility theory approach, Harding and Lovenheim (2017) show that nutrient-specific taxes have a larger impact on food choice than product-specific taxes, because nutrients are found across many foods and substituting away from the taxed nutrient (such as sugar) becomes harder for consumers.

The most important empirical regularity in consumption is the law of demand, whereby smaller quantities are consumed at higher prices, indicating a reduction in utility derived from consumption. Exceptions that prove the rule are situations where some other factor intervenes, so that higher prices lead to increased consumption. These exceptions can be Giffen goods (Jensen and Miller, 2008), which are necessities at the bottom of a preference ranking for lower-income people so that price increases drive increased consumption, or Veblen goods (Currid-Halkett, Lee and Painter, 2018), which are luxuries for conspicuous consumption that is visible to others in a particular social group, such as dining at high-cost restaurants. Both Giffen and Veblen goods can be found in the food system but account for a small share of total consumption. For almost all estimated demand systems, consumption follows the law of demand, and to the extent that revealed preference captures consumer welfare, wellbeing is lower when prices are higher.

### _Estimating demand systems_

Estimating demand systems is challenging due to the many factors that might modify each relationship. In some settings, circumstances might stay stable enough for a demand curve between price and quantity to be visible in two dimensions. For example, seasonal fluctuations in avocado prices and quantities sold in California trace out a demand curve that can be seen to shift from year to year (Huntington-Klein 2021). More complete analyses would require a multivariate system of simultaneous equations. For example, expenditure might first be allocated between different kinds of goods and services such as food, housing, transportation, and clothing, and then allocated within each category for example among food groups such as starchy staples, legumes, meat, eggs, dairy, vegetables, and fruits, leading ultimately to selection among substitutes within each food group. In each kind of demand system, we are interested first in how consumers' overall income and total expenditure drives change, followed by effects of price changes and other factors.

*Income*

The quantity consumed of an item as income rises is known as an Engel curve, derived by tracing out consumers' expansion path in expenditure at a given set of prices and other determinants of food demand. Engel's law states that food as a whole is generally a normal good with an income elasticity of demand between 0 and 1, meaning that spending rises with income but less than proportionally as other goods enter into total expenditure. Some foods are inferior, meaning that their income elasticity of demand is negative, and some are luxuries, meaning that they take a larger share of total expenditure at higher incomes so their income elasticity of demand is greater than one. A given food such as rice may be a luxury for the poorest, a normal good intermediate income levels and then an inferior good at higher incomes, as people switch between food sources to meet their various needs.

Income effects may be estimated from cross-sectional variation in quantities consumed at different income levels, as well as time-series variation in income, randomly assigned transfers, and hypothetical choice experiments. For a current review and update on behavioral and experimental approaches to agricultural economics research, readers should refer to Palm-Forster (2022). Because consumers with different income levels face similar prices, many observations of income effects are cross-sectional, as in Gouel and Guimbard (2019). Cross-sectional patterns differ from changes over time due to many unobservable factors associated with income that shift demand in persistent ways, limiting short-run income response. In both the short and long run, one of the most important patterns is that demand for total dietary energy is roughly constant, in the sense of calories per day over a given period of time, while income effects greatly alter the composition of food expenditure.

Modern studies of energy balance find that calorie intake per day is roughly constant over time because it is closely tied to a factors that change relatively slowly within a narrow range, such as a person's height, weight, body composition and physical activity level (Pontzer et al. 2021). These factors often vary with income, leading to dramatic differences in calories per person at different income levels (Logan 2009).  For an individual person, however, total energy intake is regulated by involuntary hormonal signals and physiological mechanisms that help maintain energy balance within biological limits. Controlled trials have revealed the extent of variation associated with metabolism, physical activity and weight loss or gain, as well as the influence of food composition on appetite and speed of eating (Hall et al. 2019). Energy balance can also be influenced to some degree by individual genetics (Loos and Yeo 2021) and variation in gut health and the microbiome (Lee, Sears and Maruthur 2020). These physiological constraints ensure that correlations between a person's income and their total energy intake is embodied primarily in heights and weights, with total food consumption and quantities purchased also affected by variation in kitchen and plate waste (Wilson et al. 2017, Barrera and Hertel, 2021).

The low income elasticity of demand for calorie intake is among the oldest known stylized facts in the economics of nutrition. In *The Wealth of Nations*, Adam Smith observes that "The rich man consumes no more food than his poor neighbour. In quality it may be very different… but in quantity it is very nearly the same", which Smith explains by saying that "The desire of food is limited in every man by the narrow capacity of the human stomach" while "conveniences and ornaments of building, dress, equipage, and household furniture seems to have no limit." (Smith 1776). This insight that food quantities are biologically constrained predates any understanding of metabolism; it was not until the 1780s that Antoine Lavoisier developed the ice calorimeter to measure heat associated with respiration, from which he observed that food contains energy in quantities that differ from the food's volume or weight but are related to body size and physical activity (Hartley 1947, West 2013).

The energy content of food is a scientific fact but is not experienced directly or intuitively known. A recent survey in Malawi asked respondents a set of questions designed to elicit what they knew about the foods they use, for example whether there was more energy for work each day in a glass of water or a glass of milk; salaried nurses understood that question and a large majority (83%) correctly chose milk, but for villagers the question was not meaningful and more than half (57%) chose water (Schneider and Masters 2019). Even among highly educated consumers for whom caloric content is measured and labeled, most people do not notice or remember how much energy is in the items they chose, as shown for example among restaurant customers in the United States by Cawley, Susskind and Willage (2020).

People maintain energy balance without knowing how much energy is in their food, implying that appetite must be regulated by unconscious processes. Changes in intake associated with income are not due to a conscious preference for higher energy intake, but to other factors including shifts in heights and weights, illness and gut health or physical activity levels. A change in circumstances that causes additional intake would typically manifest itself in weight gain, but could potentially be used for higher physical activity, and even drive higher income if productivity had been constrained by low energy intake. Increased calorie intake might raise productivity and incomes for very low-income people in strenuous occupations, such as farmers in India who have lower labor productivity and incomes when Ramadan fasting coincides with labor-intensive periods in cropping calendar (Schofield 2020), or rickshaw drivers in Chennai who had higher productivity and incomes when randomly assigned a high-calorie snack plus cash rather than the control group who received only cash (Schofield 2014). More commonly, correlations between a person's energy intake and their income arises due to shifts in other factors, including the use of heights or weights as social signals for discrimination in the markets for labor and capital (Clément, Levasseur and Seetahul 2020, Macchi 2021).

The unconscious nature of dietary energy regulation contributes to the difficulty of dietary recall, which may be subject to systematic biases that affect estimated income elasticities (Burrows et al. 2019). One pioneering study in a low-income setting was Bouis and Haddad (1992), whose work in the Philippines to improve survey measurement reduced their estimate of the calorie-income elasticity from 0.30-0.59 to 0.08-0.14, after addressing issues such as discarded food and measurement errors. Researchers can also measure changes in energy balance through changes in body weight, finding small changes in response to even a complete removal of any income or price constraint which occurs each school year when students join unlimited meal plans in college cafeterias (Crombie et al. 2009).

Measurement error can have a substantial impact on food demand system estimation and agricultural data collection, and readers interested in a thorough treatment of these issues should refer to Carletto, Dillon, and Zezza (2021) in this Handbook volume 5. Using a random assignment of different survey instruments compared with a gold-standard instrument, Gibson et al. (2014) find that food consumption measured with error is negatively correlated with true food consumption, especially for households in rural areas. In an introductory article to a special issue in *Food Policy* on measurement in food consumption, Zezza et al. (2017) argue that survey instruments and methods impact estimates of food consumption, and cite evidence that data errors depend on observable household characteristics such as income levels. Survey practices that do not give appropriate attention to food acquired away from home, or use wide recall periods should be discontinued, and researchers must consider various tradeoffs in survey intensity and time needed to answer each given research question (Zezza et al. 2017). To address challenges in measurement of prices and consumption, Gibson and Kim (2019) include a direct measurement of market prices to estimate the price response, instead of having to rely on an indirect budget share estimation method.

After accounting for physiological conditions and measurement issues the income elasticity of demand for calories is near zero, but we find large income effects on food expenditure and diet composition. Some of this additional expenditure is for convenience and food safety as well as taste and socio-cultural aspirations, but one of the most consistent patterns is demand for dietary diversity between and within food groups (Clements and Si, 2018). At low income levels that diversification reflects Bennett's law, by which consumption shifts towards foods other than starchy staples as incomes rise. This pattern was first observed in national averages across countries (Bennett 1941), and has been confirmed countless times in microeconomic data (Ali et al. 2018; Almas et al. 2019). That pattern can help people meet essential nutrient requirements for lifelong health (McCullough et al. 2022), but can also include diversification into newly introduced foods that may have harmful attributes (de Brauw and Herskowitz 2020). Demand for diversification also reflects Engel's law, by which expenditure diversifies from the most basic to additional human needs as income and productivity rises (Baffes and Etienne 2016). Focusing on Africa where initial food expenditure and diet diversity are low, one recent meta-analysis of 66 studies found an average income elasticity of food expenditure of 0.61 (Colen et al. 2018), a result that appears robust despite apparent publication bias in favor of larger elasticity estimates (Ogundari and Abdulai 2013). Studies of short-term income shocks such as the Asian financial crisis of 1997-98 reinforce the idea that the income elasticity of demand for total energy is low, but changes in diet quality can significantly alter intake of micronutrients (Skoufias et al. 2012, Block et al. 2004).

Economics research on income elasticities includes the composition of calories, and changes in the proportion of energy obtained from carbohydrate, fat and protein. Early work on income-calorie relationships such as Bouis and Haddad (1992), Subramanian and Deaton (1996), and Dawson and Tiffin (1998) has more recently been extended using a variety of new techniques. For example, a randomized cash transfer intervention in Kenya revealed that protein was a luxury good in that context, with an income elasticity of 1.29 (Almas et al. 2019). In other settings, the most income-elastic macronutrient is fats (Salois, Tiffin and Bacombe 2012). These patterns are non-homothetic, as only the initial stages of diet diversification away from the least-cost starchy staple towards more expensive foods introduces different nutrients that are found, for example, in animal sourced foods. In the very low-income setting of Somalia, demand for all foods except cereals and oils can be income-elastic (Hussein et al. 2021), and a meta-analysis of 90 studies found that the threshold at which income elasticities of nutrient demand declined was about the World Bank's poverty line which was $1.25/day at the time (Zhou and Yu 2015).

An important aspect of how income affects food demand is that income from different sources may be allocated to different purposes. Much of this literature from low- and middle-income countries concerns gender differences in how women and men spend their earnings, when women are responsible for meal preparation and child nutrition leading them to spend more of any additional income on higher diet quality (Quisumbing and Doss 2021). Other studies link income sources to expenditure through non-separability of markets outside the household, as in panel data from Tanzania where income elasticities of food consumption are higher for agricultural earnings than for non-farm income (Van den Broeck et al. 2021). These differences in how income from different sources is typically spent reflects differences in gender norms and the frequency at which income is earned, ranging across and within households from small and frequent payments to larger and infrequent lump sums (Valluri et al. 2021).

_Prices_
Price response is closely related to income elasticities, with own- and cross-price elasticities typically estimated along with income response using demand system models such as the linear expenditure system (Stone 1954) and the Rotterdam model (Barten 1964; Theil 1965), or translog demand systems (Christensen et al. 1975) and the Almost Ideal Demand System (Deaton and Muellbauer 1980) and its extensions such as Quadratic-AIDS and Linear Approximate-AIDS among others. For a thorough review of various demand systems models and their applications to nutrition questions, readers should refer to the textbook from Babu et al. (2016).

The responsiveness of food demand in the context of price shocks is particularly relevant for child nutritional status, given that children's growth and functioning are more sensitive to changes in intake over short periods. Vellakkal et al. (2015) find that a 10 rupee increase in the price of rice per kilogram led to a decrease in child rice consumption by 73 grams per day and a slight decrease in weight-for-height z-scores in India. Yamauchi and Larsen (2019) demonstrate that the effects of the 2007/8 food price crisis can last over the long-term, finding that child growth was negatively affected by price increases for households that did not grow their own food. Arndt et al. (2016) find that incidence of wasting and underweight decreases when food price inflation decreases. While Yamauchi and Larsen (2019) did not find evidence of gender differences in food price shocks in Indonesia, Kumar and Quisumbing (2013) demonstrated that female-headed households in Ethiopia were more vulnerable to food price shocks during the 2007/8 crisis.

Estimating both income and price elasticities is particularly important to identify how transfer programs, taxes or subsidies and other interventions alter dietary choices. For example, sugar-sweetened beverages and other low-cost sources of "empty calories" may be consumed more often by lower income people, driving regressive effects on income distribution of Pigovian taxes to discourage their use.  But lower income people may also have a higher price elasticity of demand for soda and other such products, as found for example in Mexico by Colchero et al. (2015), implying a greater health benefit from these Pigovian taxes. The fact that lower-income people may bear a greater economic burden but also obtain greater health benefits from Pigovian taxes underscores the value of targeting revenue to those who consumed the product prior to the intervention, including harm reduction programs for improved outcomes among those who continue to consume the harmful product despite the tax.

The choice of which demand system to use in any particular setting depends on several factors. For example, what is the research question or the main parameter to be estimated? What data are available to do the estimation? And what contextual issues are most important for external validity of the estimates? In terms of food demand, in some cases there will be a high degree of substitutability between goods, and in other cases there will not be. In agricultural economics, the AIDS and its extensions are preferred because of its flexible functional form, its ability to aggregate over consumers, and its ease of interpretation (Alston and Chalfant 1993; Cornelsen et al. 2014).

Some researchers have estimated demand for nutrients derived from food consumption, aiming to test for alignment between derived demand and the production of human health. For example, Huang (1996) transformed a food demand system into derived demand for nutrient intake given the fixed nutrient composition of 35 foods and their demand elasticities. Subsequent work has sought to address challenges of model specification and measurement, including contextual factors that vary by season and location (Hoddinott and Stifel 2019). In rapidly changing food systems it may be especially important to use updated data, as estimates drawn from the literature may not reflect current conditions (Chen et al. 2016).

Repeated estimation of food price elasticities in different situations permits meta-analysis such as Green et al. (2013), Cornelsen et al. (2014), and Chen et al. (2016). These typically find that income and price elasticities of demand are lower at higher incomes, and that cross-price elasticities vary greatly across studies. Clements and Si (2016) combine previous studies to calculate compensated price elasticities after accounting for income changes. The uncompensated price elasticity of demand is the percentage change in the consumption of a particular food item when the price changes by a given percentage. Changes in the choice of consumption bundle involve the substitution effect and the income effect, and the uncompensated change in demand for a good after a price change is equal to the difference between the substitution and income effects. The substitution effect occurs when the consumer changes the mix of goods purchased due to relative price shifts between the goods in the bundle. The income effect occurs when a price increase for a good has caused real consumer income to decrease, and therefore the entire consumption bundle is affected.

The compensated price elasticity of demand focuses on pure substitution effects, supposing hypothetically that consumers are compensated for the loss of income from the price change to keep utility constant. However, since relative prices between goods have changed, consumers still substitute between goods and the responsiveness of those changes can be measured with the compensated elasticity of demand. Marshallian (uncompensated) demand functions, where the quantity consumed of a good is a function of prices and income, holds income constant and measures the changes in quantity consumed due to price. Hicksian (compensated) demand function maps the quantity consumed as a function of prices and utility, so utility is held constant in the elasticity estimates while income is allowed to change. Hicksian demand is estimated by minimizing consumer expenditure, whereas Marshallian demand is estimated by maximizing consumer utility.

Changes in uncompensated demand can be related to changes in compensated demand using the Slutsky identity, in which the uncompensated price elasticity of demand for good *A* equals the compensated price elasticity of demand for good *A* (the pure substitution effect) minus the income elasticity of good *A* times the expenditure share of good *A* (the income effect). This identity also pertains to cross-price elasticities, in which the uncompensated price elasticity of demand for good *A* with respect to a change in the price of good *B* equals the compensated cross-price (pure substitution) elasticity minus the income elasticity of good *A* times the expenditure share of good *B*. Therefore, when trying to understand demand changes for foods, four main factors matter: the own-price elasticity of the food, the substitution effect between foods when relative prices change, the income effect, and the budget share of the good in question.

The full effect of a price change on quantity demanded depends on the net change of the combined substitution and income effects. The decomposition of price changes is highly relevant for the economics of nutrition. First, in terms of human health there are many potential ways to obtain sufficient nutrients for supporting life, therefore understanding compensated substitution between goods is essential. Second, the income effect interacts with price changes in complex ways, especially for farm families who are producers as well as consumers and may be either net sellers (who benefit from a price rise) or net buyers (who are harmed by it), and for analyses that span households with very different budget shares for food. For all these reasons, keeping track of both price and income effects is helpful for understanding changes in food consumption and diet quality.

*Information and willingness to pay*
Information asymmetry can drive high-quality items out of the marketplace (Akerlof 1970), and lack of information more generally can have a variety of market-distorting effects and consumer welfare implications. Providing more accurate information is typically helpful but can have unintended consequences. For example, Li et al. (2020) examine the consequences of a change in the nutrient profile algorithm used by third party *NuVal* software to characterize the healthiness of foods, and found that producers responded to anticipated demand changes by raising the prices of items whose score improved. Edenbrandt et al. (2018) use a hedonic price model to demonstrate that consumers value the information from food package labels. While labeling for nutritional quality by a third party may have some short-term benefits for individual consumers, those benefits may be counterbalanced by the potential for worsening price discrimination, the introduction of less healthy products, and increased market segmentation, especially when it comes to packaged and processed foods.

Not all health-related information is accurate, and people may use health information in surprising ways. Companies that market their products as a fun indulgence often avoid good-for-you labeling, and when they do reformulate products to be healthier they may reveal that information selectively so that only health-oriented buyers would notice (Lacy-Nichols et al. 2020). Furthermore, observed correlations between adherence to dietary recommendations and health outcomes could reflect other health-seeking behaviors, some of which may be unobservable. Oster (2020) shows how consumers whose dietary intake changed in response to changing nutrition guidance are likely to take other health-promoting actions as well, exaggerating the estimated impact of improved diet quality alone. Finally, consumer concerns about specific diseases may affect purchasing decisions, as demonstrated by Thapaliya et al. (2017) in the case of direct-to-consumer markets.

The development of new foods and supplements, or new labeling or marketing arrangements, often involves market trials and choice experiments to estimate of consumer willingness to pay (WTP). For example, a Becker-DeGroot-Marschak random-price auction experiment in Ghana with women seeking pre- and post-natal care found that 69.4 percent of participants had a WTP higher than the production cost of $1.33 (2011 USD) for a particular nutritional supplement, and those WTP estimates vary significantly with pregnancy or breastfeeding status, household wealth, and previous experiences with an infant with malnutrition (Adams et al. 2016). Context-specific WTP estimates of this type have high utility for informing the design of nutrition policies and programs, for example to show differences in demand for micronutrient-fortified products in urban versus rural areas of Kenya (Pambo et al. 2017).

### *Education, empowerment, and the use of information to guide consumption*
The use of information to guide consumption is mediated by education and empowerment, provided for example by more and better schooling, employment, legal rights and cultural or social protections. Education and empowerment enable people to act on what they already know, and has been found especially important to improve maternal and child nutrition (Webb and Block 2004). Much nutrition education consists of advice about what to eat, but schooling also leads to very different knowledge about basic facts such which foods contain more energy and whether orange soda has the same composition as orange fruit (Schneider and Masters 2019). Education also leads to employment that changes the flow of income to a household, shifts intrahousehold bargaining dynamics among family members, and changes the opportunity cost of a mother's time (Quisumbing and Doss 2022). All these factors may affect child nutrition in either positive or negative ways, depending on the mechanism that dominates (Alderman and Headey 2017).

Maternal employment might raise tradeoffs in time use and resource allocation that affect child nutrition, and overall effects of labor force participation is likely to vary with contextual factors. In Egypt, Rashad and Sharaf (2019) found that maternal employment was negatively associated with child nutritional status when estimated using an IV strategy with maternal employment instrumented by local labor market conditions. In Nepal, women's empowerment is positively associated with maternal nutritional status and child height-for-age (Malapit et al. 2015). As the nutrition transition continues, understanding the linkages between maternal employment and child overweight and BMI will be more important as well. Current evidence in low- and middle-income countries suggests that maternal employment is not associated with child overweight but may improve BMIs that are too low in poorer countries (Oddo et al. 2017).

*Retail prices and the affordability of least-cost diets for nutrient adequacy and lifelong health*
Most economics research on food demand focuses on observed consumption, but some studies focus on food environments and access to healthy diets. Research on diet costs and affordability aims to determine the extent to which retail prices of least-cost healthy foods, relative to available income, lead to malnutrition simply because healthy diets are out of reach and cannot be acquired without a change in available foods, prices or income. Modeled diets that meet nutritional criteria at least cost can inform antipoverty programs and safety nets, as in the Thrifty Food Plan that shows how recipients of nutrition assistance can meet all their dietary requirements within the legislated benefit level (USDA 2021), or the cross-country work of Allen (2017) who showed that the World Bank's $1.90 poverty line is typically insufficient for people to meet their nutrient needs.

In recent years, high speed computation of least-cost diets using retail prices of many diverse items each month at each market location has been used to reveal spatial and temporal variation in access to nutritious diets, as in Omiat and Shively (2017) and then Masters et al. (2018). These studies focused first on least-cost foods to achieve nutrient adequacy, and then extended to the least-cost items in each food group needed to meet national dietary guidelines as in Herforth et al. (2020). This approach has been adopted by UN agencies as a new metric of food access, focused on the central finding that an estimated 3 billion people could not afford a healthy diet in the 2017-19 period (FAO, IFAD, UNICEF, WFP and WHO, 2021). Quantifying food access by computing least-cost diets links the retail prices of locally available items to their nutritional characteristics, helping to guide intervention in food supplies and safety nets that could ensure universal access to healthy diets. Once healthy diets are affordable, other factors might drive the choice of items actually consumed including constraints on time use and meal preparation, food preferences and many other factors that could change with economic development.

## 2. The nutrition transition and economic development
Observed changes in dietary intake and nutritional status associated with economic development are driven by the interaction of price and income with innovations in food production, distribution and marketing. Drivers of the nutrition transition include forces from outside of the food system such as urbanization and the impact of lower transport and communication costs. Within the food system, nutrition transition is closely linked to the structural transformation of agriculture, whereby farm production accounts for declining share of employment and income as economies develop over time.

The nutrition transition described by Popkin and others results from the interaction between many different changes that are also associated with economic development. These include the demographic transition from high to low birth rates associated with declining child mortality and longer lifespans; the epidemiological transition from infectious to non-communicable disease associated with vector control, immune response and public health; the structural transformation of economic activity from agriculture to services and industry associated with capital accumulation, innovation and income growth; the agricultural transformation of farming from low to high productivity through improvements in crop and livestock genetics, input use and resource management; and above all the dietary transition, defined as the shift in consumption away from farm-grown foods usually consumed at home and towards the consumption of more packaged, processed and prepared foods often consumed away from home. At the individual level, these economy-wide changes manifest as fluctuations and patterns in wealth, income, education, prices, and preferences. Work in agricultural economics focuses on how changes to income, prices, and preferences affects nutrient intake, and how various economic phenomena interfere with healthy choices, such as time discounting, asymmetric information, intrahousehold allocations, and the role of policy and regulation.

## 2.1 Structural transformation and the food system

The classic story of structural transformation, summarized in the inaugural volume of the *Handbook of Development Economics* series (Timmer 1988), was formulated using data from national accounting and employment statistics starting with Kuznets (1946) and Chenery (1960). These data were seen to follow systematic patterns long before the nutrition transition was observed, as labor and other resources flowed from agriculture towards services and industry partly through rural-urban migration, and partly through increasing specialization and nonfarm activity within rural areas that drives food system change.

Timmer (1988) describes the agricultural transformation as having four phases. In the initial stage, after centuries of roughly constant per-capita incomes, something changes to raise labor productivity and household income. Agricultural innovation may trigger the start of income growth, but since farmers have had centuries of experimentation, the change more often starts with off-farm discoveries such as improved transport or an industrial revolution. Then in the second phase, due to the attractiveness of other sectors, as well as diminishing returns on a fixed land area and inelastic demand for farm products, there is an accelerating flow of labor and capital out of agriculture into higher-return activities in services and industry.

Sustaining the second phase of agricultural transformation in isolated settings that must produce their own food requires agricultural productivity growth, otherwise rising food prices slows or stops any further growth. In open economies that can be fed through increased imports, agricultural productivity growth is needed primarily to raise farmer incomes and increase output of tradable goods. In both cases, farm productivity growth typically requires complementing farmers' own experimentation with outside inputs, from public sector investment in new technologies, infrastructure and institutions to raise farm output per unit of land and livestock.

The third phase of agricultural transformation happens as the flow of people and resources out of agriculture into the nonfarm sector gradually slows, due to increasingly close integration of rural factor and product markets with the nonagricultural sector. Earnings are initially much higher in the non-farm sector, driving movement out of agriculture until farm income rises due to either productivity increase or a rise in land area per worker. Once all available farmland has been brought into production, land area per worker depends on the number of workers, which changes with natural increase net of out-migration. The peak number of farmers in any country typically happens quite late in its agricultural transformation, for example around 1914 in the United States, after which rural out-migration allowed the remaining farmers to cultivate an increasing area of land and livestock per worker.

A fourth and perhaps final phase of agricultural transformation, seen in the U.S. since the 1990s, is when the outflow of labor from agriculture has ended and the farm sector functions similarly and alongside other economic sectors. In South Korea, the share of employment in agriculture fell from 65 percent in 1960 to 5 percent in 2018, a much faster process than in the United States or Great Britain (Teignier 2018). Farming differs from nonfarm work in many ways, but in the fourth stage of the agricultural transformation, farmers are no longer systematically much poorer than non-farmers. Each kind of farm activity has and continues to change in many ways during all phases of the transition, with many big consequences for the food system and nutrition such as increasing geographic specialization driven by location-specific comparative advantages relative to transportation costs, and other forms of specialization as some farms produce commodity inputs to food manufacturing while other farms produce increasingly differentiated food products.

The transformation of each country's agri-food system is both cause and consequence of change in dietary intake and nutritional outcomes (Webb and Block 2012). Growth in per capita income and urbanization are closely intertwined characteristics of the agricultural and structural transformations. In combination, these characteristics have direct and profound effects on food environments, including changes in tastes and dietary preferences along with modernization of agricultural marketing systems and supply chains.

## 2.2 The nutrition transition in undernourishment and obesity

To describe how economic development relates to the nutrition transition, we use two of the most widely recognized and frequently cited kinds of nutritional data: first the prevalence of undernourishment computed annually for every country by the FAO, and then the prevalence of overweight or obesity estimated annually for every country by the WHO. These two metrics both reflect energy balance from macronutrients, which is also influenced by linear growth and attained height relative to international standards. The prevalence of low height for age, known as stunting, as well as low weight for height known as wasting, are estimated most often from Demographic and Health Survey (DHS) data, which are based on a nationally representative population of mothers with children under 5 years of age in many low- and middle-income countries.

*Linear growth and stunting*
A person's height would reach their genetic potential if their linear growth from infancy through adolescence were not constrained by diet quality and disease. Faltering can occur at any stage of growth, but an important discovery of the 1990s and 2000s is that most of the world's growth faltering to a lower height trajectory actually occurs before age two (De Onis and Branca 2016). Since a child's attained height and weight is among the most easily measured kind of nutritional status, and since many kinds of malnutrition and illness manifest as linear growth deficits, the heights of children under five have long served as the single best marker of a population's overall nutritional status. Short stature, relative to a reference population of healthy children, is associated with many adverse health and socioeconomic outcomes later in life such as reduced cognitive functioning, lower schooling, and increased morbidity and mortality. Early intervention can help prevent these effects, leading to higher educational attainment and higher wages in adulthood (Behrman, Hoddinott, and Maluccio 2020).

Anthropometric data to measure linear growth is typically transformed by converting absolute lengths into percentiles or standard deviations (z scores) of the distribution in a healthy reference population. The definition of stunting is having a height-for-weight z score less than 2 standard deviations below the median at their age and sex, so a healthy population would have about 2.5% of its children below that threshold at any given time due to individual variation in genetic potential and the timing of growth spurts. Any individual's height at one point in time should not be used to diagnose their health, but repeated measurements in childhood are highly informative (Cliffer et al. 2021). Sustained and widespread faltering, for which high stunting prevalence is the most severe manifestation, is a sign of impairments that are likely to have affected many aspects of child development. Short stature as such can be an adaptation to poor diets, without which undernourished children might be too tall for their circumstances and not have survived at all (Perumal et al. 2018), thereby providing one of the most useful available markers of adverse exposures in early life (Leroy and Frongillo 2019).

The economics literature on stunting addresses its causes and consequences for health, productivity and well-being, and the evaluation of programs to help children fulfill their genetic potential for linear growth and other aspects of child development. Research on the causes and consequences of stunting can be framed in terms of debate around Vollmer et al. (2014), who used successive DHS surveys to show that changes in national income had limited correlation with changes in stunting. Some of the correlation that is observed relates to diet composition, such as Headey et al. (2018) regarding use of animal sourced foods in infant feeding to provide the high density of bioavailable nutrients needed for child development. RCTs have demonstrated that complementary feeding promotion for children aged 6 to 24 months can improve HAZ and WAZ (Bhutta et al. 2013). Nutritional supplements and related strategies such as reducing malaria incidence and improving access to sanitation can also improve child growth (Keats et al. 2021). Empirical evidence on the effectiveness of nutrition interventions is improving, but is still challenged by unmeasured confounding and measurement error and would benefit from more longitudinal studies in diverse contexts, such as the Institute of Nutrition of Central America and Panama (INCAP) work in Guatemala (Behrman et al. 2020. Many of the linkages are long-term in nature, however, such as the consequences of child malnutrition for adult productivity and income. For example, Galasso and Wagstaff (2019) use a development accounting approach to estimate that the GDP lost from deprivations associated with stunting are 5 to 7 percent per capita, and 9 percent in areas with higher burdens of linear growth faltering. Other estimates are higher, with Horton and Steckel (2013) estimating that GDP losses were 12 percent in low-income countries due to chronic undernutrition, proxied by attained male height. In survey data, a typical finding is that each centimeter of additional height is associated with 6 percent higher wages for women, and 4 percent higher wages for men (McGovern et al. 2017).

*Ponderal growth, wasting and overweight or obesity*
Weight gain trajectories have very different dynamics from linear growth, driven by fat deposition and muscle development around a person's bone structure. Maternal malnutrition can lead to poor gestational health and low birthweights, after which poor infant feeding can lead to low weight-for-length that interacts with linear growth faltering (Thurstans et al. 2021). With exclusive breastfeeding, children can typically fulfill their genetic potential for weight and height up to about six months of age, after which the gradual introduction of diverse solid foods to complement breastmilk would be needed but is often not provided (Choudhury et al 2019). Insufficient diets (in both quantity and quality) and disease with or without breastfeeding can lead to extreme thinness or wasting, and the onset of wasting may be quite sudden. Once detected, clinical or at-home treatment can lead to complete recovery to a normal weight trajectory within a few weeks, however recovery may not be sustained even with the added intervention of antibiotics and nutrition counseling (Lelijveld et al. 2021). Wasting often goes untreated due to lack of access to care, and episodes of wasting over time leads to linear growth faltering as well (Stobaugh et al. 2018).

Low ponderal growth, for which the threshold of wasting in children is defined as having a weight-for-height more than 2 SD below the median of a healthy population, or underweight in adults as having a BMI below 18.5 kg/m$^2$, typically involves both low adiposity and limited muscle development. In contrast, excess weight gain towards overweight and obesity, defined as having a BMI above 25 and 30 kg/m$^2$ respectively, typically reflects gradual fat deposition and associated changes in metabolism. BMI and its thresholds offer an imperfect metric of body composition but can help detect unwanted weight gain for individuals and groups. As with other health conditions, early detection of unwanted gain is helpful due to the asymmetry between fat deposition and weight loss, as physiological mechanisms favoring homeostasis work to sustain body mass once it is formed. Many different factors could trigger the initial occurrence of unwanted weight gain, including the introduction of new highly processed foods that lead to faster intake of more dietary energy than would be chosen with more traditional, bulkier foods (Hall et al. 2019).

*Measuring changes to macronutrient intake*

Dietary energy from macronutrients is the most fundamental aspect of food consumption, driving metabolism and body composition over the life course. The first macronutrient requirements to be identified involve proteins, so insufficient intake has long been known as protein-energy malnutrition. More recent research reveals an equally important role for dietary fats in macronutrient balance (Mozaffarian, Rosenberg and Uauy 2018). Given the difficulty of direct measurement through dietary recall, global monitoring of total dietary energy intake from all macronutrients is done using the prevalence of undernourishment (PoU) metric developed by Sukhatme (1961) for the FAO to track whether global agricultural production and food supplies were keeping up with population growth in the 1960s. A population's PoU is the estimated number or fraction of people for whom the available food supply provides less total dietary energy than a healthy and active population would require.

The FAO's global undernourishment metric is updated annually and provides a headline number through which to track the energy adequacy of national food supplies and distribution (FAO, IFAD, UNICEF, WFP and WHO, 2021), complementing the limited number and frequency of household or individual surveys,. To compute the PoU for each year, the country's population pyramid of people at each age and sex is combined with an estimate of their estimated minimum dietary energy requirements if they had a healthy body size and physical activity level. That total energy requirement is then compared to the total dietary energy available in the country, estimated in food balance sheets from national total food production plus imports minus exports and estimated loss or non-food uses for each agricultural commodity. The distribution of that food supply is then allocated using a lognormal function parameterized with past household consumption and expenditure surveys, and used to estimate the prevalence of undernourishment in each year (Cafiero, 2014).

Figure 1 presents the scatterplot and nonparametric relationship between the prevalence of undernourished population and log Gross National Income (GNI) per capita, along with regionally differentiated country-level data points (e.g., a Preston curve for undernutrition). With this level of country aggregation, undernutrition declines monotonically with per capita income over time and across countries. There is a wide range of variation around the norm, but countries with PoU levels above the mean at each income level typically converge towards that mean over time. These patterns are driven by differences and change in countries' estimated total food supply, the age structure of their population, and the lognormal distribution function estimated from their household consumption surveys.

**Figure 1. Prevalence of undernourishment at each level of national income**

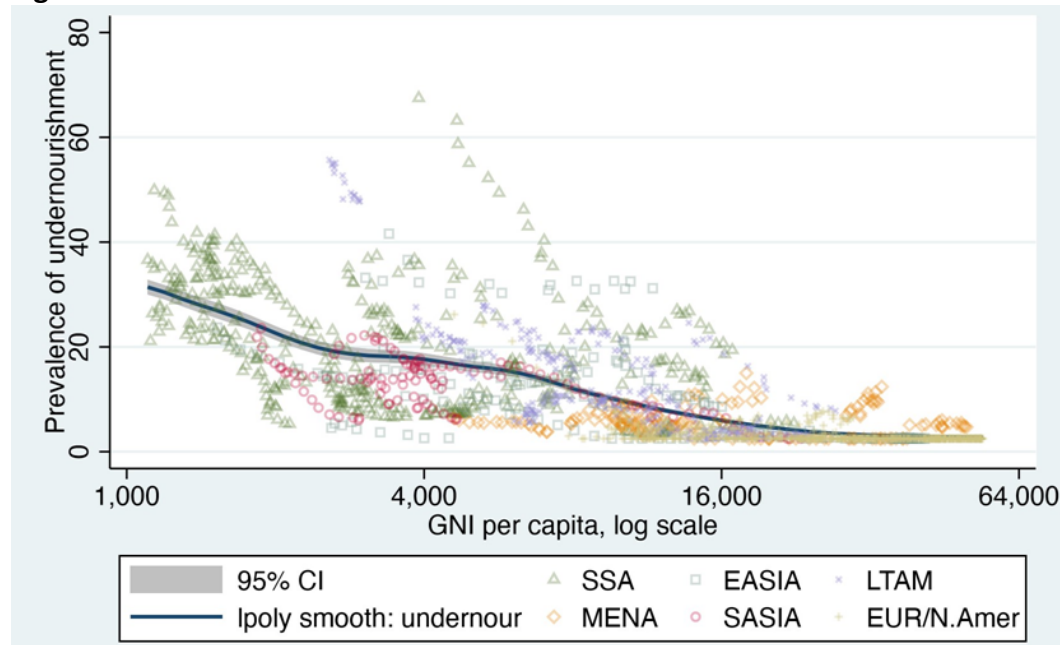

Figure 2 presents the same kind of data visualization for the fraction of adults in each country and year whose BMI exceeds a reference threshold of 25 kg/m². Estimates are computed for the World Health Organization using a Bayesian hierarchical model from available surveys, smoothed across countries and over time (Abarca-Gómez et al. 2017). Figure 2 demonstrates that the second step of nutrition transition, in which overweight and obesity become increasingly prevalent, differs from the decline in undernourishment in several important ways. First, global mean at each income level rises nonlinearly through an S-shaped curve, most steeply in the range of about $3,000 to $10,000 per year in 2017 US dollars. Second, there is little apparent convergence to the global mean, as countries above or below that line do not move towards it over time. These differences could be artifacts of the estimation methods, but could also be driven by empirical differences between the two concepts, as the prevalence of overweight can begin to be addressed as soon as policymakers take steps to prevent unintended excess weight gain, and when individuals have access to evidence-based nutrition therapy and other supportive resources.

**Figure 2. Prevalence of overweight and obesity at each level of national income**

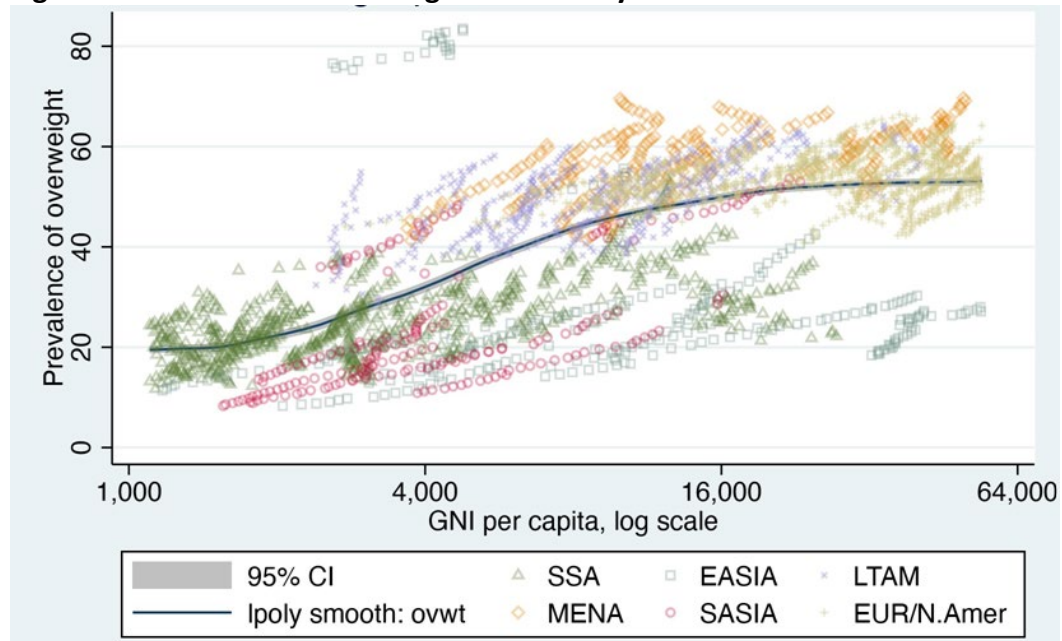

Note: Data shown are WHO estimates from the World Development Indicators (World Bank, 2021), for 217 countries from 1990 to 2017, at each level of gross national income per capita in terms of 2017 U.S. dollars at purchasing power parity prices. The prevalence of overweight or obesity is defined as the percentage of a country's population whose weight and height exceeds 25 kg/m2. Estimates are modeled using a Bayesian hierarchical model from available surveys, smoothed across countries and over time. The solid line is a local polynomial regression with its 95% confidence interval for the global mean at each income, computed using Stata's lpolyci with default bandwidth.

As shown in Figures 1 and 2, the nutrition transition in undernourishment and overweight or obesity differs greatly among countries and continents. Collapsing the visualization into a stylized picture over just the two dimensions in Figure 3 reveals how continents differ in the degree of double burden, meaning the prevalence of overweight at each level of undernourishment.

Figure 3 is an adaptation of the classic "Ruttanogram" that shows transition from increasing yields per acre to increasing output per worker during the agricultural transformation (Hayami and Ruttan 1985). Diagonal lines show the level of double burden, in the sense of having more malnutrition in both dimensions. A path that traces a 45° line upwards towards the left would indicate that undernutrition declined at the same rate that overweight increased. Regions for which the path crosses a 45° line from below saw overweight increase more rapidly than undernutrition declined, and vice versa for Africa which is the only region to have experienced a decline below the line where 50% of the population is either undernourished or overweight. Across regions, Latin America and the Caribbean comes closest to tracing a path along a 45° line.

**Figure 3. Nutrition transition and the double burden of undernourishment and overweight**

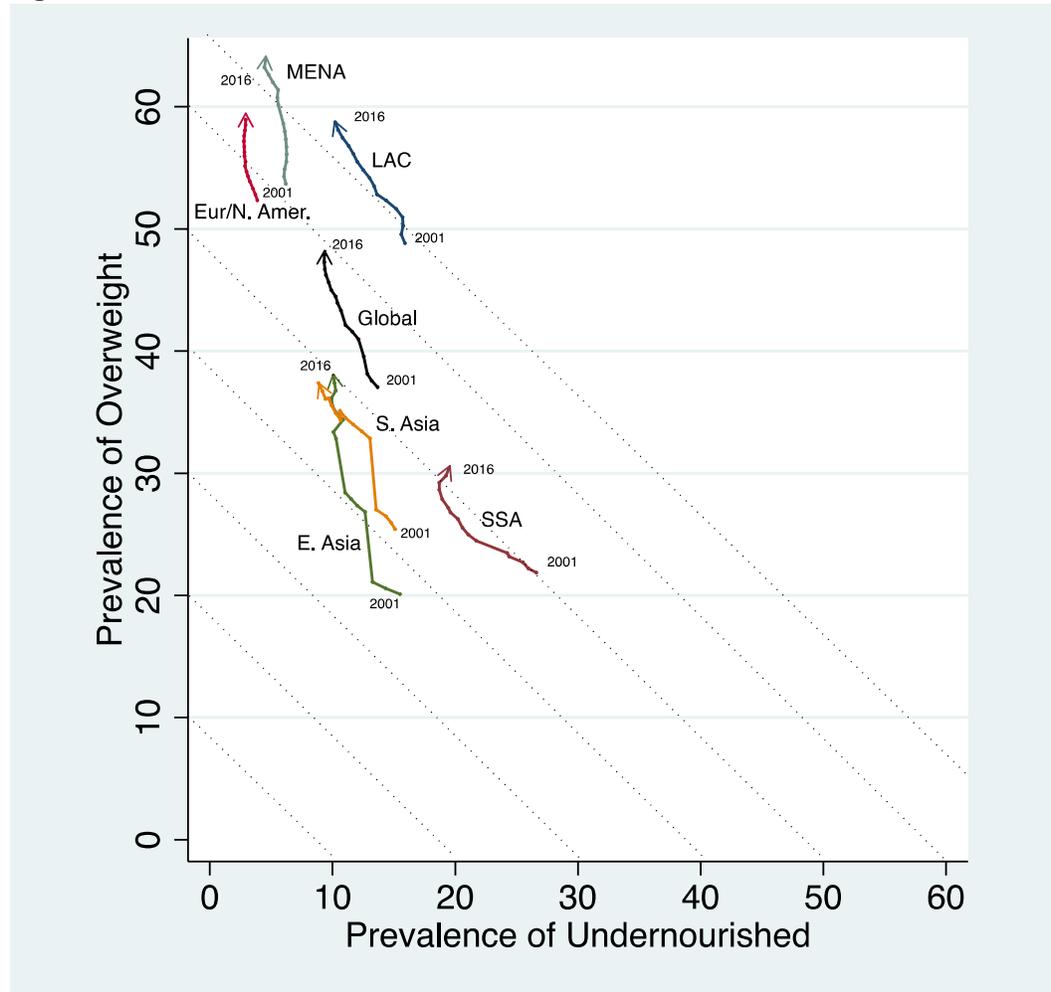

Note: Data shown are from Figures A and B, collapsed to their mean for each continent in each year in the 21[st] century, from 2001 to 2016.

Figure 3 clearly reveals that most regions and the world as a whole are characterized by much more rapid increases in overweight than their rates of progress in reducing undernutrition. The only exception is Sub-Saharan Africa, yet even that region reversed course towards the end of this period, crossing a 45° line from below with an increase in the double burden during a period with increases in both overweight and undernourishment. It is notable, as well, that by 2016 South Asia had surpassed East Asia in reducing the prevalence of undernourishment, although as noted earlier these modeled estimates include artifacts of the methods as well as empirical differences in the underlying observations.

Figures 1, 2 and 3 show the prevalence of undernourishment and overweight as trajectories of each country or region over time, at each level of per capita national income shown in Figures 1 and 2. These income trajectories involve many kinds of structural change, including especially the agricultural transformation associated with increasing urbanization whose links to undernourishment and overweight are shown in Figures 4 and 5. In each of these two figures, Panel A shows national trajectories for the bivariate relationship itself, and Panel B shows that same relationship after controlling for national income (in log form). The vertical axis for Figure 4 shows each country's prevalence of undernourishment and Figure 5 shows its prevalence of overweight or obesity. The horizontal axes show each country's urban population share, with the red lines in each Panel A showing the nonparametric smoothed relationship and in Panel B showing the semiparametric relationship after controlling for income.

Figure 4 reveals that the prevalence of underweight is closely associated with urbanization when considered two dimensionally in Panel A, but when controlling for income in Panel B the relationship disappears, except for time trends shown by the trajectories of individual countries. This observation strongly suggests that developments other than urbanization as such are what drive down the prevalence of underweight in each country over time.

**Figure 4. Prevalence of undernourishment at each level of urbanization**
Panel A                                                    Panel B

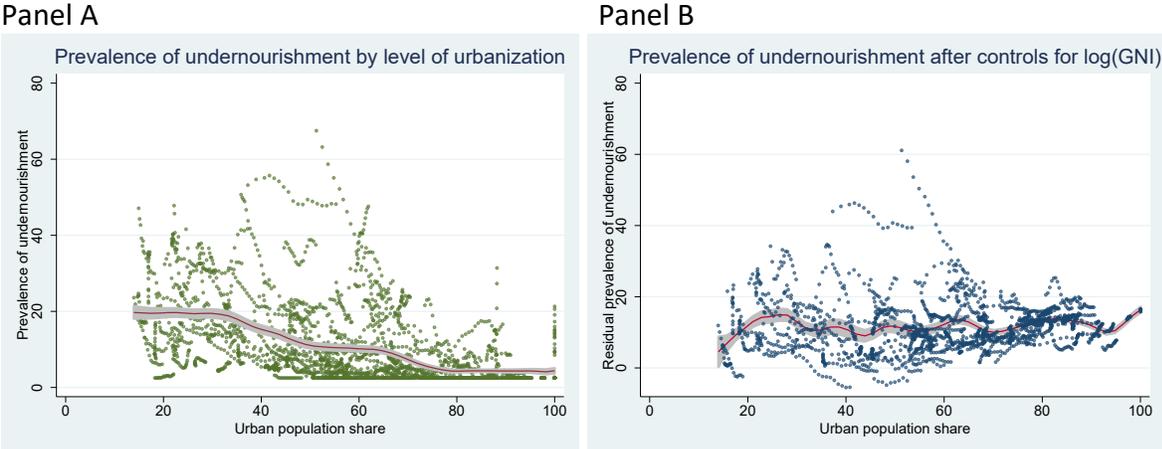

Note: Data shown are FAO estimates from the World Development Indicators (World Bank, 2021), for 127 countries from 2001 to 2016, at each level of urban population share. The prevalence of undernourishment is defined as the percentage of each country's population for whom available dietary energy is below the minimum required for a healthy and active person of their age and sex. Estimates are modeled using UN population data, food balance sheets and a distribution function for the food supply in each country and year. The solid line is a local polynomial regression with its 95% confidence interval for the global mean at each income, computed using Stata's lpolyci with default bandwidth.

Figure 5 repeats this comparison of non-parametric and semi-parametric associations for the prevalence of overweight. Again, the two dimensional relationship in Panel A shows a close association with urbanization. In striking contrast to Figure 4, however, controlling for income in Panel B makes no difference to that relationship. While factors associated national income other than urbanization appear to drive down undernutrition, the rise of overweight and obesity appears to be driven by factors more closely tied to urbanization as such rather than per capita income. The individual country trajectories show rising overweight at each level of urbanization whether or not the scatterplot controls for income, again suggesting that rising overweight and obesity are due to structural factors in the food environment that change over time rather than income itself.

**Figure 5. Prevalence of overweight at each level of urbanization**

Panel A

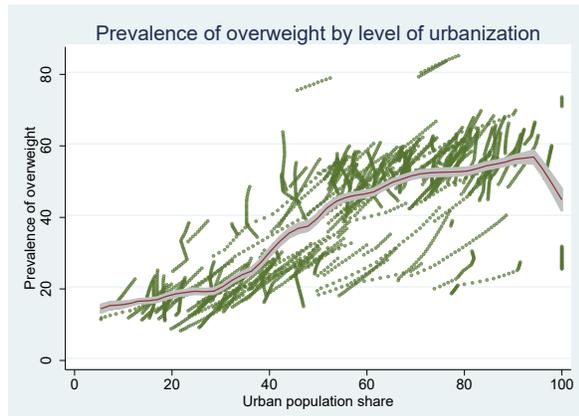

Panel B

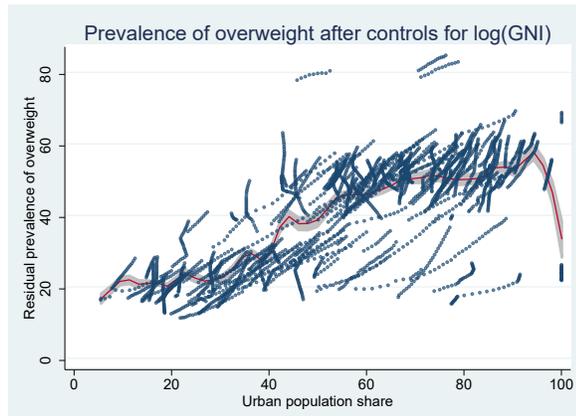

Note: Data shown are WHO estimates from the World Development Indicators (World Bank, 2021), for 147 countries from 1990 to 2016, at each level of urban population share. The prevalence of overweight or obesity is defined as the percentage of each country's population whose weight and height exceeds 25 kg/m2. Estimates are modeled using a Bayesian hierarchical model from available surveys, smoothed across countries and over time. The solid line is a local polynomial regression with its 95% confidence interval for the global mean at each income, computed using Stata's lpolyci with default bandwidth.

Greater insight into shifts in the prevalence of undernourishment and overweight are shown in Figures 6 and 7, which provide the cross-sectional frequency distribution of each prevalence at at roughly ten-year intervals, in 1995, 2005, and then 2016 (the most recent available year), for each of six major world regions.

In Figure 6 we see that Europe and North America, and to lesser extent the Middle East and North Africa, began with low levels of underweight and experience little change over these years. At the other end of the scale, Sub-Saharan Africa and South Asia began with the highest regional mean levels of underweight. While both regions saw significant reductions between 1995 and 2005, Sub-Saharan Africa then stagnated while South Asia saw continued reductions in its regional mean prevalence of underweight through 2016. In East Asia, the most salient change over these years was the narrowing distribution across countries, rather than a rapid reduction in the regional mean, suggesting catch-up by the initially more prevalent countries.

**Figure 6. Frequency distribution of undernourishment by region in 1995, 2005 and 2016**

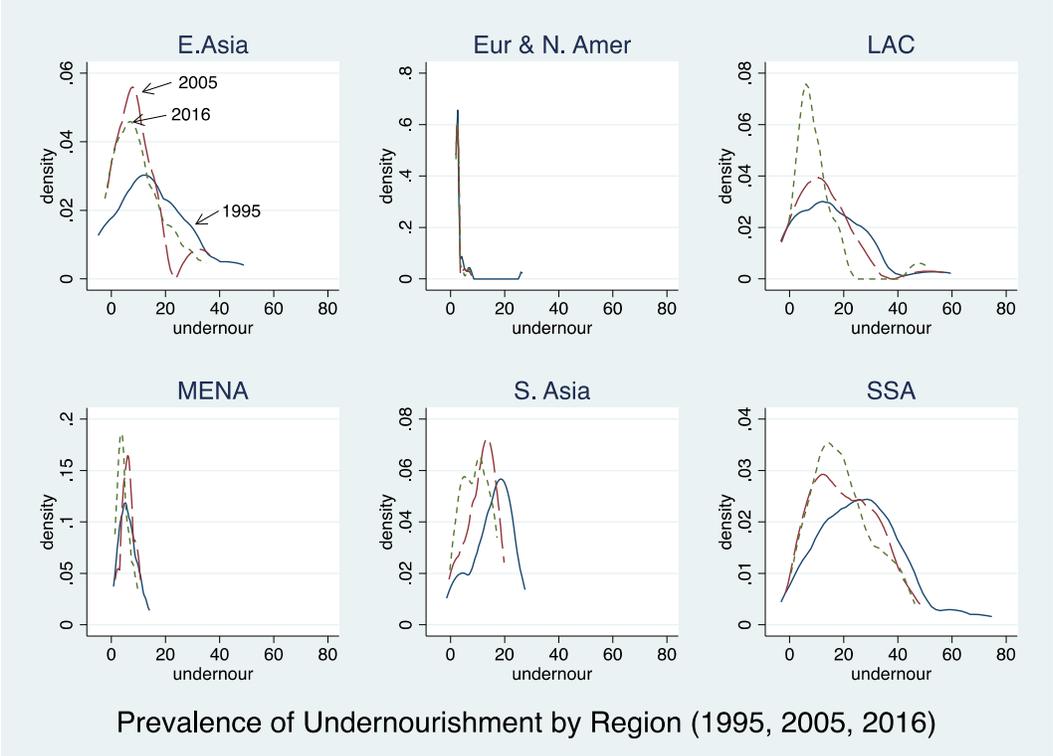

Prevalence of Undernourishment by Region (1995, 2005, 2016)

Note: Non-parametric kernel density functions of the prevalence of undernourishment by region, defined as the percentage of each country's population for whom available dietary energy is below the minimum required for a healthy and active person of their age and sex.

The march towards increasing prevalence of overweight forms a much more robust pattern across regions. Figure 7 shows an increasing country mean prevalence in every region, albeit from differing initial levels and at differing rates of change. While East Asia, South Asia, and Sub-Saharan Africa all experienced increasing mean levels of overweight between 1995, 2005, and 2016, these regions' initial levels were relatively low and increased relatively slowly. In contrast, the Middle East and North Africa, Latin America and the Caribbean, and Europe and North America all began with prevalence of overweight levels on the order of 40 percent and all three regions saw rapid increases to over 60 percent prevalence by 2016.

**Figure 7. Frequency distribution of overweight by region in 1995, 2005 and 2016**

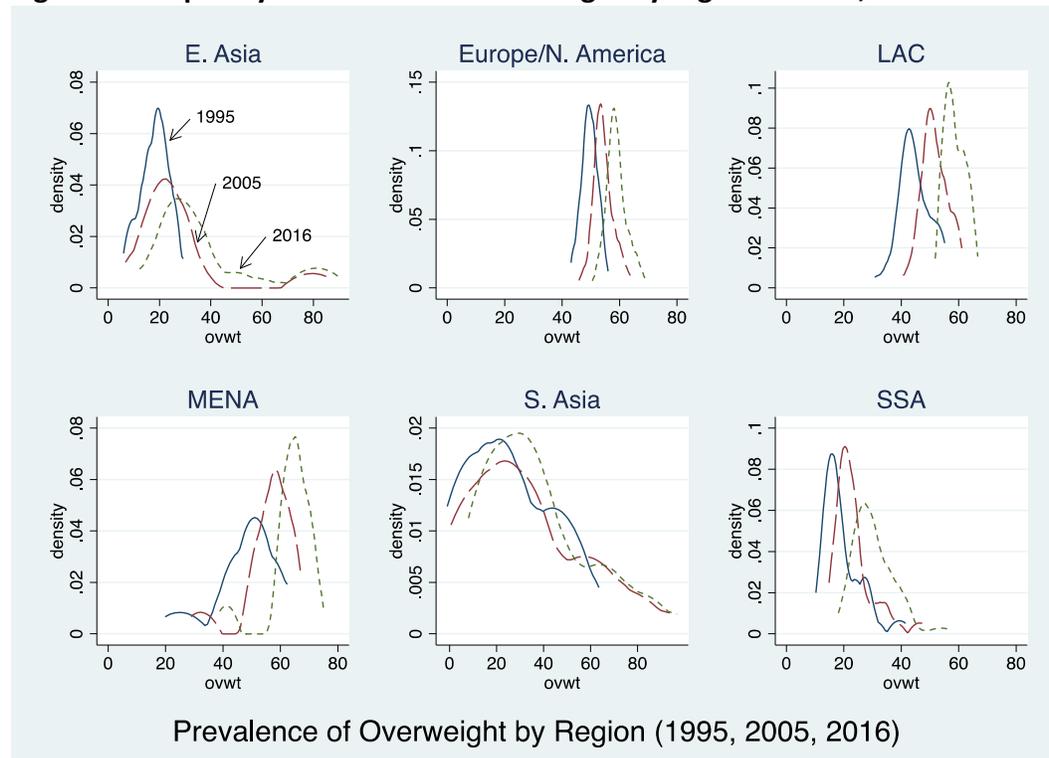

Prevalence of Overweight by Region (1995, 2005, 2016)

Note: Non-parametric kernel density functions of the prevalence of overweight or obesity by region, defined as the percentage of each country's population whose weight exceeds 25 kg/m2.

## 2.3 The nutrition transition in micronutrient deficiencies

A third aspect of the triple burden of malnutrition is micronutrient deficiencies, and insufficient or excessive intake of other compounds known to affect health. Inadequate intake of essential vitamins and minerals remains a major public health concern in low- and middle-income countries, as gaps are often filled at higher incomes by dietary diversification, fortification or supplementation. Beyond the 20 or so micronutrients that are widely monitored by health authorities and added to foods and dietary supplements for many decades, malnutrition can also involve imbalanced consumption of countless more recently discovered and still to-be-discovered compounds including the type of fat, protein and carbohydrates that provide macronutrient energy, antinutrients that affect absorption and use of other food components, fiber and other pre- or probiotics that affect gut health and microbiota, as well as antioxidants and other compounds involved in disease processes (Barabási, Menichetti and Loscalzo 2020).

To illustrate the economic aspects of micronutrient deficiency we trace the dietary transition in one vitamin and one mineral that are particularly important nutrients of concern: first iron, which is most often deficient for adolescent girls and women due to its role in hemoglobin and blood health, and then vitamin A which is most often deficient when diets contain insufficient quantities of animal source foods (e.g., liver, fish, eggs), and dark green, orange, or red fruits and vegetables and animal-source foods. These and other micronutrient deficiencies have historically been most influenced by changes in the mix of foods consumed, but once each compound is identified, deficiencies can be addressed directly by four kinds of intervention: (1) industrial fortification or enrichment of processed food, (2) biofortification through traditional and transgenic and fertilization of crops or feed to animals to alter micronutrient levels in raw foods, or (3) supplementation of the diet with additional micronutrients delivered in tablets, drops, powders or injections, and (4) demand generation in the sense of nutrition education or assistance programs to promote more diverse diets. Fortification and enrichment as well as biofortification affect the entire food supply, while supplementation and demand generation programs can be more precisely targeted to at-risk individuals and households.

Our two example micronutrients, iron and vitamin A, are chosen in part to illustrate how the nutritional biochemistry affects the economics of deficiency prevention and treatment. Iron is a water-soluble mineral for which frequent intake is necessary, as relatively little can be stored as ferritin in the liver and elsewhere. In contrast, vitamin A is fat-soluble, and several months' supply can be stored in the liver. Furthermore, iron absorption depends greatly on the form and context within which the mineral is found in the diet, whereas vitamin A can more readily be metabolized and absorbed from both animal and plant sources. In both cases, nutritional requirements are somewhat but not always reflected dietary preferences.

Micronutrients generally cannot be seen, tasted or smelled, but are present in raw foods in approximately fixed proportions, and the role of those foods in lifelong health might have shaped taste preferences and culinary norms long before biochemical research, randomized trials and epidemiological studies identified which compounds affect what outcomes. Animal-source foods that are rich in heme iron and vitamin A, as well as vegetal foods that are rich in non-heme iron and vitamin A, have long been seen as desirable components of a balanced diet, but these foods are often expensive and revealed preferences are unlikely to fully reflect what more recent research has discovered about how these or other micronutrients affect future health. For that reason, completing the nutrition transition towards healthy diets often requires deliberate intervention to improve micronutrient levels through targeted fortification and supplementation, as well as dietary change in the diversity of foods consumed.

*Iron deficiency and anemia*
Iron deficiency is the most common micronutrient deficiency worldwide, leading to adverse health and cognitive outcomes including effects of iron deficiency anemia (IDA) in which insufficient blood hemoglobin limits oxygen transport needed for metabolism. Iron deficiency anemia can be treated with oral supplementation or increasing intake of iron-rich and fortified foods, as well as intravenous infusion and blood transfusions in extreme cases. Prevention of iron deficiency is possible through promoting dietary diversity and supplementation among vulnerable groups especially during growth and development or menstruation.

Considerable efforts have been made to increase the availability of iron in foods, especially beans, through fortification or biofortification (Anderson 2020). A key factor for improving iron status is in ensuring the bioavailability of iron in diets, which is a challenge because most plant foods have high levels of anti-nutrients which bind to iron and reduce bioavailability (Anderson 2020). In addition to fortification and biofortification, iron deficiency can be prevented with iron and folic acid supplements (IFA), or multiple micronutrient supplements (MMS) for pregnant women (Bourassa et al. 2019). These supplements are cost-effective and reduce the risk of adverse birth outcomes (Bourassa et al. 2019). For treating IDA among pregnant women, intravenous iron sucrose is effective at increasing maternal iron stores (Bhavi and Jaju 2017).

Iron deficiency is one of several interacting factors affecting the incidence, severity and duration of anemia, which is also linked to vitamin B-12 and folate as well as blood loss and parasitic diseases such as malaria, exposure to lead and other heavy metals, or some chronic diseases. Anemia symptoms of economic importance include fatigue, headache, dizziness, and difficulty thinking or concentrating, all of which contribute to lower educational attainment, productivity and earnings (Chaparro and Suchdev 2019). Economics research about iron deficiency addresses not only its causes and consequences, but also program evaluation for prevention and treatment. For example, an observational study in India found that women's empowerment is associated with higher intake and less severe deficiency (Gupta, Pingali and Pinstrup-Andersen 2019), a randomized controlled trial in rural China of a nutrition information program for parents found that increasing knowledge about anemia reduced the risk of anemia in children by 6.1 percent (Zhao and Yu 2020), and a trial of iron-fortified salt for school meals in India reduced anemia prevalence in children by 20 percent (Krämer, Kumar and Vollmer 2021).

Figure 8 illustrates regional trends in the prevalence of anemia among women of reproductive age and children under 5. While the global trend is towards diminishing prevalence, rates remain particularly high in Sub-Saharan Africa and South Asia. It is of additional concern in the former case that the prevalence of anemia among children is substantially greater than the prevalence among women, as iron deficiency in the first 1000 days is strongly associated with long-term impairment of cognitive development.

**Figure 8. Prevalence of anemia by region, 1990-2015**

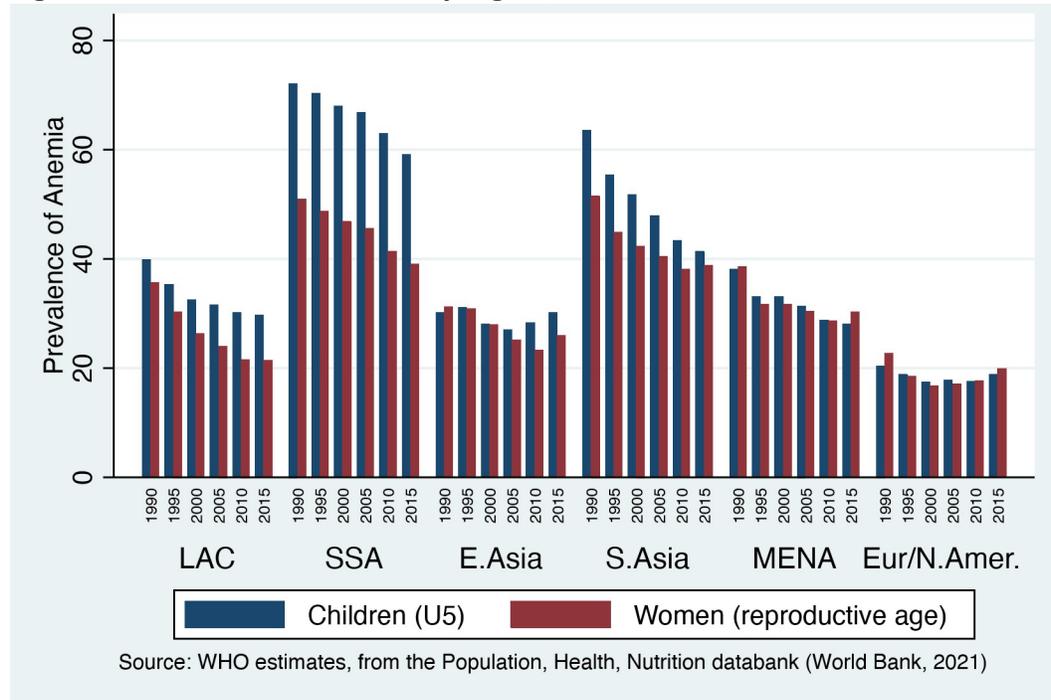

Source: WHO estimates, from the Population, Health, Nutrition databank (World Bank, 2021)

Note: Data are WHO estimates from World Development Indicators (World Bank, 2021) for 139 countries for years 1990-2016. Prevalence of anemia among children ages 6-59 months is the percentage of children whose hemoglobin level is less than 110 grams per liter, adjusted for altitude. Prevalence among non-pregnant women is the percentage whose hemoglobin is less than 120 grams per liter at sea level.

Figures 9 and 10 demonstrate the strong inverse relationship between prevalence of anemia and income per capita.

**Figure 9. Prevalence of anemia in children under 5 at each level of national income**

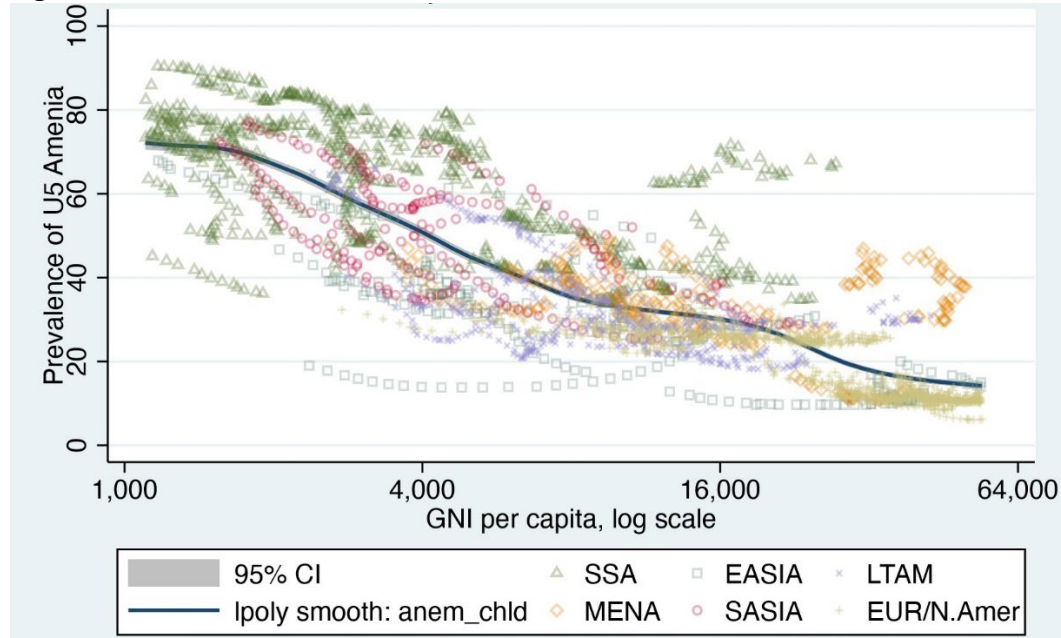

Note: Data are WHO estimates from World Development Indicators (World Bank, 2021) for for 139 countries for years 1990-2016. Prevalence of anemia among children ages 6-59 months is the percentage of children whose hemoglobin level is less than 110 grams per liter, adjusted for altitude, as a function of gross national income per capita in terms of 2017 U.S. dollars at purchasing power parity prices. The solid line is a local polynomial regression with its 95% confidence interval for the global mean at each income, computed using Stata's lpolyci with default bandwidth.

**Figure 10. Prevalence of anemia in women at each level of national income**

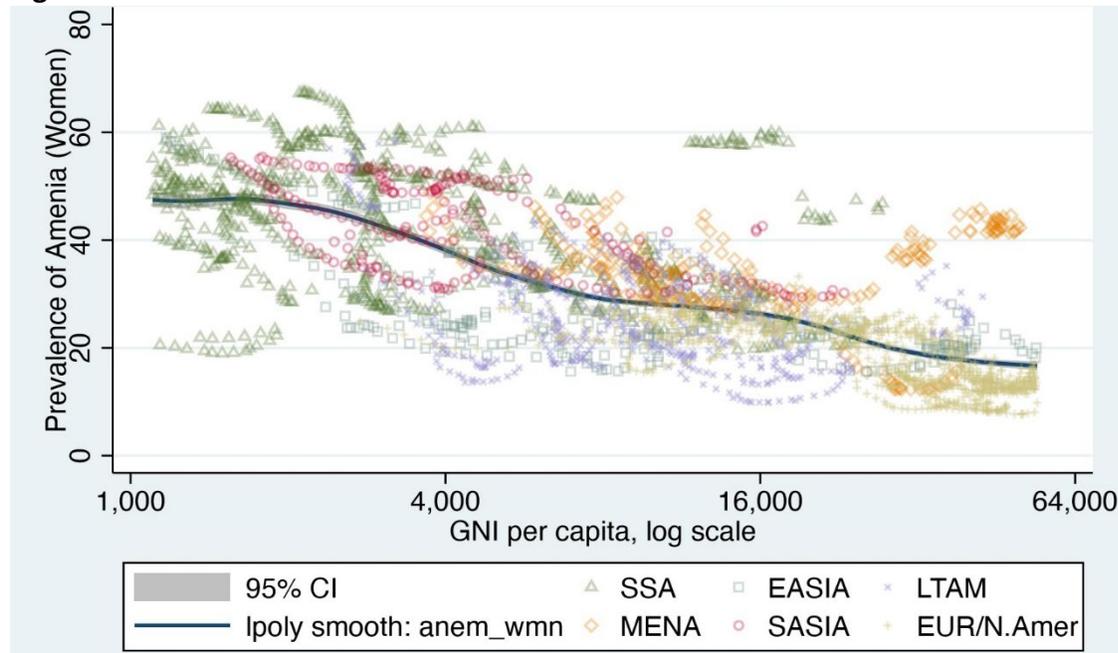

Note: Data are WHO estimates from World Development Indicators (World Bank, 2021) for for 139 countries for years 1990-2016. Prevalence of anemia among non-pregnant women is the percentage whose hemoglobin level is less than 120 grams per liter, at sea level, as a function of gross national income per capita in terms of 2017 U.S. dollars at purchasing power parity prices. The solid line is a local polynomial regression with its 95% confidence interval for the global mean at each income, computed using Stata's lpolyci with default bandwidth.

*Vitamin A deficiency and its consequences*

The richest sources of vitamin A are the retinoids stored in animal sourced foods especially liver, fish, and eggs, while plant carotenoids metabolized into vitamin A are concentrated in dark green and yellow-orange fruits and vegetables. Many of the fruits and vegetables that are rich in provitamin A are perishable and seasonally produced, contributing to the difficulty of consuming enough to avoid deficiency in the off-season. Vitamin A is essential for growth and development, cell differentiation, vision, and immune function. Deficiency of Vitamin A is the leading cause of preventable blindness in the world, and increases the risk of mortality from infectious disease, particularly measles and diarrheal disease for children (Jones and de Brauw 2015).

Figure 11 demonstrates a negative correlation between countries' prevalence of VAD and GNI per capita. This link is similar to that of iron deficiency, but nonparametric regression reveals that most rapid decline in the prevalence of VAD occurs approximately between $2000 and $8000 levels of income per capita (const. 2017 $PPP). The relatively flatter association above $8000 per capita appears to be driven by the persistently higher rates among the middle-income countries of Sub-Saharan Africa.

**Figure 11. Prevalence of vitamin A deficiency at each level of national income**

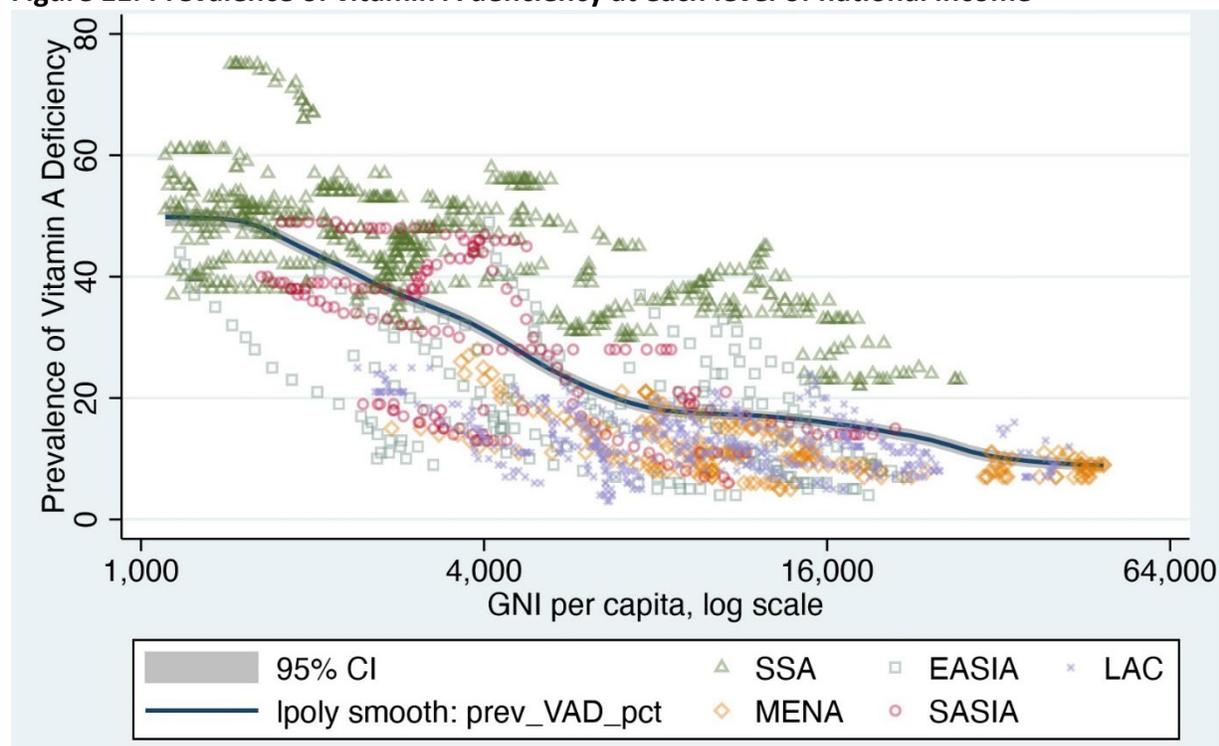

Note: Data on Vitamin A deficiency are from Stevens et al. (2015). Vitamin A deficiency is defined as a serum retinol concentration lower than 0.70 $\mu$mol/L, modeled 134 population-representative data sources for 83 countries for years 1991-2013. Data on gross national income per capita are in 2017 U.S. dollars at purchasing power parity prices. The solid line is a local polynomial regression with its 95% confidence interval for the global mean at each income, computed using Stata's lpolyci with default bandwidth.

Given the serious consequences of vitamin A deficiency and the feasibility of achieving adequacy through supplementation, many countries now include biannual megadoses in their child health services. Additional work in agricultural and development economics is focused on raising dietary intake through vitamin A-rich foods and also biofortification of crops (De Brauw et al. 2018; Lividini et al. 2015), and iron-rich beans (Oparinde et al. 2016). Lividini et al. (2015) find that widespread adoption of a biofortified pro-vitamin A rich maize variety in Zambia could reduce the total disease burden by 23 percent over 30 years, measured with Disability Adjusted Life Years (DALYs), with a cost per DALY averted is $24. Work on dietary diversification towards orange and dark green vegetables or fruits is also important, such as promotion of orange-flesh sweet potato (OFSP) (De Brauw et al. 2018; Jones & De Brauw 2015). In Mozambique, promoting OFSP in a community-randomized intervention reduced child morbidity from diarrheal disease. Caeiro and Vicente (2019) find that a nutrition education program about OFSP and vitamin A increased and sustained nutrition knowledge as well as promoted the cultivation of OFSP by female farmers in Mozambique, although these changes did not come with increased measured OFSP consumption by participants.

De Brauw et al. (2018) found that improvements to dietary intake of vitamin A in Mozambique and Uganda after OFSP promotion were mainly due to the adoption of the crop by farmers and not due to improved nutrition knowledge. A choice experiment in the field in Uganda found little difference in consumers' willingness to pay for biofortified OFSP with or without a promotional campaign (Chowdhury et al. 2011). The most important characteristic of OFSP for consumers appears to be its density and similarity of the eating experience to traditional foods rather than vitamin content (Naico and Lusk 2010). The fact that carotenoids in vitamin A-rich plants tend to color each food product helps signal its nutrient density but can also be an obstacle to adoption, as where orange or yellow biofortified maize is being introduced to populations that typically prefer white maize flour (De Groote et al. 2011; Muzhingi et al. 2008).

To identify whether and when consumers might be willing to consume more nutrient-rich foods, quantitative choice experiments and qualitative focus groups can elicit consumer preferences for key attributes of health-promoting foods, to inform policy and promotional campaigns (Jefferson-Moore et al. 2014). When individual or community-level randomization is not possible, analyses can rely on matching techniques or other strategies for causal inference with observational data. For vitamin A research in agricultural economics, this includes the analysis of national interventions like the Integrated Child Development Scheme (ICDS) in India, which provides in-kind meals and take-home rations. Mittal and Meenakshi (2019) analyze the ICDS noting that the impact of the program depends on intrahousehold dynamics, preferences for target foods (i.e., whether the transfer is infra-marginal or extra-marginal for a given household), and the income elasticity of foods. They find that the effects of the ICDS on diet quality depend on the age of the participating child, but that overall there were positive impacts on total energy but not vitamin A intake.

Beyond supplementation, biofortification and diet diversification towards vitamin A-rich foods, it is also feasible to add preformed vitamin A to cereal flour, sugar or other packaged foods. As shown by Bobrek et al. (2021), industrial fortification of wheat and maize flour is much more common for iron and folate than for vitamin A and other micronutrients. Sugar fortification is more controversial due to concerns about the harmful consequences of promoting sugar as a beneficial food, but has been examined for example in Kenya where consumers are willing to pay between 77% and 300% premiums for fortified sugar depending on their age group and life stage (Pambo et al. 2017).

## 3. Dietary transition and food choice

The cross-country patterns and temporal trends in nutrition described above are closely related to dietary patterns and food choice. The stylized facts of food system transition start with Bennett's Law: as incomes rise, consumption shifts away from unprocessed starchy staples towards more expensive sources of daily energy including animal source foods, packaged and processed foods, and foods prepared outside the home. The food system transition associated with higher income includes a rise in kitchen and plate waste as well as larger body sizes and higher total calorie intake per person (Barrera and Hertel 2021). Like other transitions associated with economic development, these trends are linked to individual household income but also systemic change in the food environment.

**3.1 Change over time in global dietary patterns**

The global transition in dietary patterns by food group is best described using FAO food balance sheet data, estimated from national data for every country in the world since 1961. Food balance sheet data show each country's estimated food used for human consumption, computed from total production of each food, plus imports minus exports, nonfood uses, and losses prior to acquisition by households. These data are known to differ from dietary recall surveys (Del Gobbo et al. 2015), but modeling efforts to estimate long-term global trends from the few available surveys yield conflicting results (Beal et al. 2021). Even if there were surveys for each country every year, they would need to have very large sample sizes to cover the entire population, and would need to be conducted year-round to account for seasonality of diets, so it is likely that nutrition researchers will continue to complement surveys with national accounts used for food balance sheets like the data shown in Figure 12.

Panel A of Figure 12 is designed to show the entire dietary transition for select regions and the world as a whole, tracing changes from 1961 to 2018 in total dietary energy per person along the horizontal axis, and in diet diversification away from starchy staples along the vertical axis. In 1961, the region with the lowest estimated dietary energy and least diet diversity is East Asia. That region's food consumption shifted rightward from under 1800 to about 2500 kcal/person before rising from about 27% to 35% of calories from sources other than starchy staples, then rising diagonally to a level in 2018 that is similar to Europe in 1961, with the same level of dietary diversification and a higher level of total dietary energy than the whole world average in 2018. Southeast Asia and South Asia also converged towards the world average, while Africa moved more horizontally. These horizontal movements are closely linked to demographic changes in age, heights and weights of the population, as well as the magnitude of food waste as shown by Barrera and Hertel (2021).

Vertical differences in Panel A reflect diversification away from starchy staples, and both variables usually but do not always increase over time. For example, Europe's changes from 1961 to 1998 could be seen as an extension of the world average trend over this period, but from 1998 to 2018 there was no further diversification and relatively small oscillations in total calories per day. Meanwhile the United States has much higher use of foods other than starchy staples, and its total dietary energy expanded sharply from 1961 to 2005 before falling back to 2018. These data are estimated with considerable error so small differences could be statistical artifacts, but the general patterns are very likely to reflect actual shifts in food used for human consumption, not just dietary intake but also efforts at food waste reduction.

Panel B of Figure 12 disaggregates the vertical movements from Panel A into calories per person from each major food group, for the world as a whole. The most dramatic change characterizing the dietary transition is rapid rise in vegetable oils, more than doubling from under 120 to over 280 kcal/person over this period. Meat and offals use doubled, from the same level to 240 kcal/person. Sugar and sweeteners rose sharply from 1961 until 1984, while eggs and dairy have risen sharply only after 1994. Fruits, vegetables, and fish and seafood have risen more steadily throughout the period, while pulses declined from 1961 to 1993 before stabilizing and rising slightly after 2008. These patterns vary by region, of course, but provide a clear and straightforward picture of what did and did not change greatly in the global food system in recent decades. These same data have been used to show how the distribution of foods and nutrients has become more equal over time (Bell et al., 2021).

**Figure 12. Dietary transition to more food and different foods, 1961-2018**

Panel A. Dietary energy per person, and percent from sources other than starchy staples

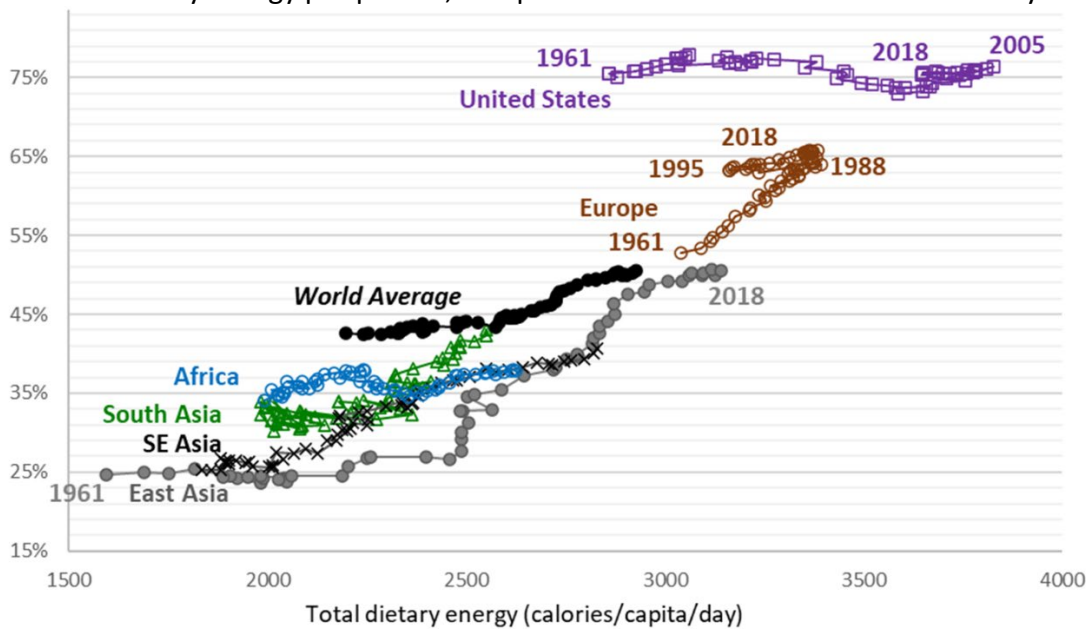

Panel B. Dietary energy per person from food groups other than starchy staples (kcal/day)

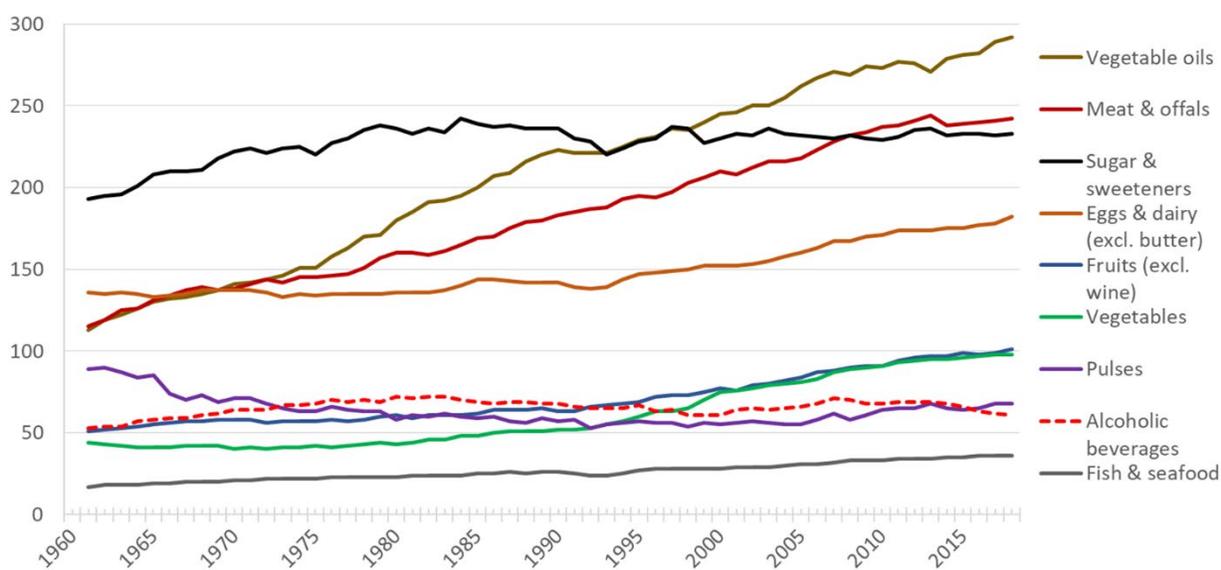

Note: Data shown are FAO Food Balance Sheet estimates of total food supply consumed by people each year (national production plus imports minus exports, non-food uses and estimated waste outside the home), merging data using new methods for 2014-2018 with earlier data for 1961-2013, all downloaded 3 April 2021.

**3.2 Change over time in U.S. dietary patterns**

A more detailed view of later stages in the dietary transition is available using national food availability data by for the United States, compiled by the USDA for 1970 through 2017. To examine post-1970 trends within this one country these are preferable to the FAO food balance sheets primarily due to a greater investment in country-specific nutrient composition data matching food industry reporting of product sales in each food group, and also a greater investment in estimating food loss at every stage of distribution including kitchen and plate waste at the point of consumption (USDA ERS, 2021).

**Figure 13. Dietary energy per person from selected foods in the U.S., 1970-2017**

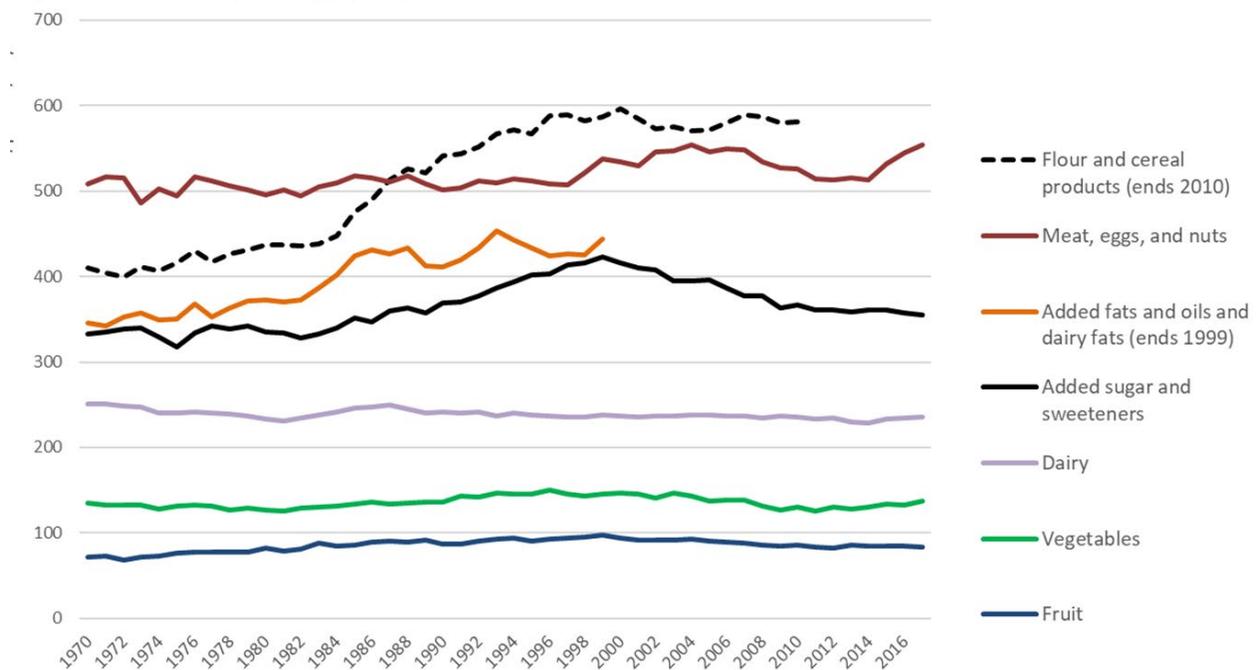

Note: Data shown are USDA estimates of loss-adjusted consumption per capita of each major food group, not shown after 1999 for added fats, and after 2010 for flour and cereals, due to changes in underlying data sources. Data were last updated 26 Aug. 2019 and downloaded 5 Apr. 2021, from https://www.ers.usda.gov/data-products/food-availability-per-capita-data-system.

A remarkable fact revealed by the data in Figure 13 is the abrupt rise in flour and cereals consumption from 1981 to 1996, when the increase slowed and plateaued after its peak in 2000. Added sugars also rose during this period, with the increase starting in 1982 and a steady decline after its peak in 1999. The rise in consumption of cereals (primarily wheat products) in the 1980s and early 1990s was a dramatic reversal of Bennett's Law, by which starchy staples tend to decline as a fraction of food consumption. In the U.S. these were primarily refined grains, and accompanied by the rise in added sugars, during an era in which consumers sought to replace dietary fat with other foods. Consumption of meat, eggs and nuts rose from 1997 to 2005, then fell back and rose again after 2014, while total consumption of dairy, vegetables and fruits changed relatively little over the years.

**3.3 Dietary transition towards packaged foods and food service across countries**

To quantify the dietary transition to packaged foods, the best data source is Euromonitor's proprietary estimates of sales by brand in selected countries assembled in their Passport database. To compile these data, Euromonitor consultants in each country maintain contact with food manufacturers, wholesalers and retailers to combine their reports with national government statistics.

Figures 14 and 15 use the Euromonitor Passport data reports to show country trajectories in sales of prepared packaged foods and carbonated beverages at each level of national income per capita. Per household annual sales of packaged foods increase slowly at low levels of national income per capita, and accelerate upwards at increasing rates as countries reach middle-income status. In contrast, sales of carbonated beverages increase rapidly between approximately $4000 and $8000 income per capita. Baker et al. (2020) focus more specifically on sales trends in ultra-processed foods including desserts, baked goods, ice cream, and ready meals. The quantity sold of these products per person is far larger in North America and Western Europe than in developing and emerging economies, but is growing at much faster rates in those lower-income regions. Between 2006 and 2019, quantity sold of these foods rose by 50% in Africa, over 50% in South and South-East Asia, 45% in Central and East Asia, and 27% in Latin America and the Caribbean (Baker et al. 2020).

**Figure 14. Sales (per household) of packaged food at each level of income, 2006-2019**

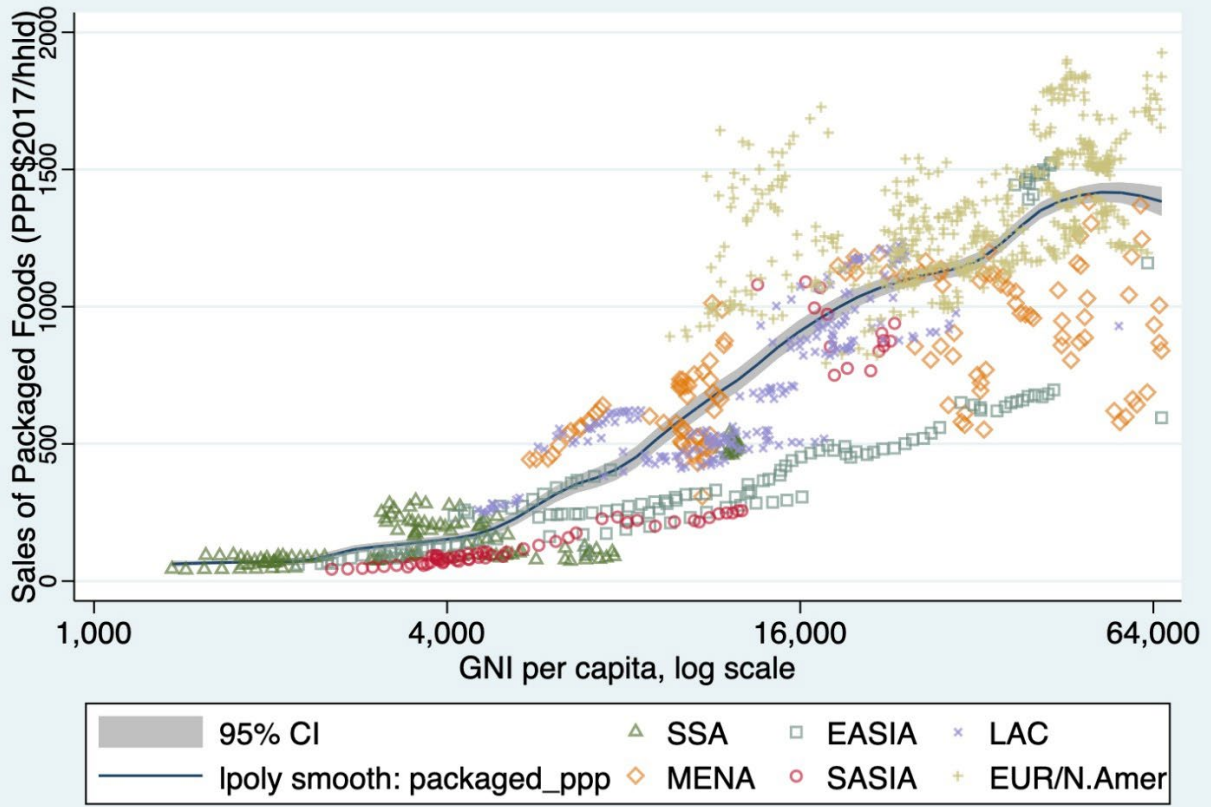

Source: Euromonitor and World Development Indicators

Note: Data shown are for 94 countries from 2006-2019, from the Euromonitor Passport database. Packaged foods data are annual sales per household. Both sales and GNI are in constant 2017 US dollars at purchasing power parity prices. The solid line is a local polynomial regression with its 95% confidence interval for the global mean at each income, computed using Stata's lpolyci with default bandwidth.

**Figure 15. Total sales of carbonated beverages at each level of income, 2006-2019**

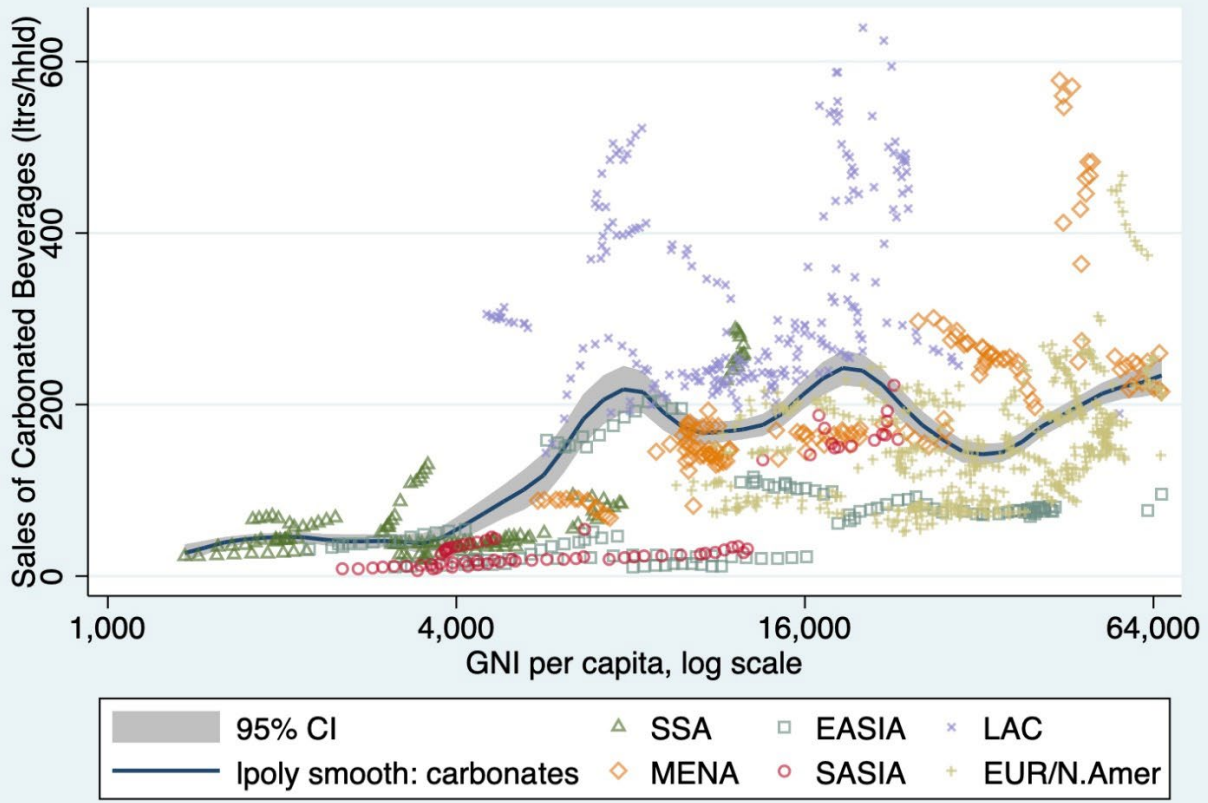

Source: Euromonitor and World Development Indicators

Note: Data shown are for 97 countries from 2006-2019, from the Euromonitor database. Carbonated beverage sales are measured in liters per household per year. Data on gross national income per capita are in 2017 U.S. dollars at purchasing power parity prices. The solid line is a local polynomial regression with its 95% confidence interval for the global mean at each income, computed using Stata's lpolyci with default bandwidth.

The nutrition transition in terms of shifts to packaged foods and items such as carbonated beverages has proceeded hand-in-hand with a profound transformation of food marketing systems and supply chains, which speaks to supply side effects on diets. Barrett et al. (2020) conceptualizes the agri-food value chain transformation as occurring in three stages. The first stage is characterized by traditional value chains. The key distinguishing feature of traditional value chains is their limited scope. Low levels of urbanization and limited physical and communications infrastructures constrain traditional value chains to local markets, thus limiting as well as the scope for intermediation. Such markets generally concentrate on starchy staples with little quality differentiation and requiring little post-harvest processing, thus limiting additional off-farm value added activities.

Barrett et al. (2020) describe a second, transitional, stage of value chain transformation. Increasing urbanization spatially stretches value chains, as the concentration of demand becomes increasingly distant from production. As suggested above, the pattern of demand from increasingly well-off urban consumers shifts towards higher-value and more perishable foods including animal sourced foods, fruits and vegetables, and processed foods. They describe the emergence of quality and food safety standards, and the flourishing of peri-urban supply chains for the most highly perishable commodities. Market intermediation deepens, as post-harvest processing and transportation services are required to accommodate the changing patterns of food demand. These services become increasingly capital-intensive, even while farming itself may remain primarily labor-intensive. They go on to describe a final, modern, stage in which agri-food value chains focus almost exclusively on urban markets, noting "The growth in urban consumer demand compels sourcing from greater distances and this increased investment in cold chains, packing, preservation, storage, bulk transport and other logistics… Consumer demand increasingly favors non-nutritive characteristics of foods." (p. 24). Supermarkets, often linked to multinational corporations, come to dominate retail and wholesale sectors, driving high levels of vertical integration. The modern stage consolidates the emergence of product standards, begun in the transitional stage in response to consumer demand safety and quality, as well as environmental and social characteristics of food production and processing.

Numerous studies focus on the type of retail outlet used for food sales, including especially the diffusion of supermarkets throughout the developing world. These studies find a variety of impacts, but their predominant conclusion is that supermarkets have had negative effects on nutrition outcomes, principally through their association with increased consumption of ultra-processed foods with harmful attributes. In a study of processed foods consumption in Guatemala, Asfaw (2011) found that a 10 percentage point increase in the budget share of highly processed foods increases BMI by 4.25 percent. Kimenju et al. (2015) found similar effects from shopping in supermarkets among adults in Central Province, Kenya. In a related study in Kenya, Rischke et al. (2015) found that supermarket purchases increased consumption of processed foods in place of unprocessed foods and increased per capita calorie intake. They note that the nutritional effects in their study are unclear, as lower cost per calorie could benefit calorie poor households while contributing to overweight in other households. This latter effect was also among the findings of Demmler et al. (2017), who join Kimenju et al. (2015) in showing a causal link between shopping in supermarkets and increased BMI. In contrast, Umberger et al. (2015) do not find a statistically significant relationship between shopping in supermarkets and adult nutrition outcomes in urban Indonesia, while Debela et al (2020) found a positive relationship with child nutrition outcomes in urban Kenya.

In parallel with the expansion of supermarkets, Barrett et al. (2017) among others describe a fast-food revolution as food consumed away from home has grown along with urbanization. Figure 16 demonstrates the strong positive association between income food service sales per capita, with a notable acceleration once countries reach approximately $20,000 income per capita. This pattern is quite similar to that observed for packaged food sales.

**Figure 16. Total restaurant and food service sales at each level of income, 2006-2019**

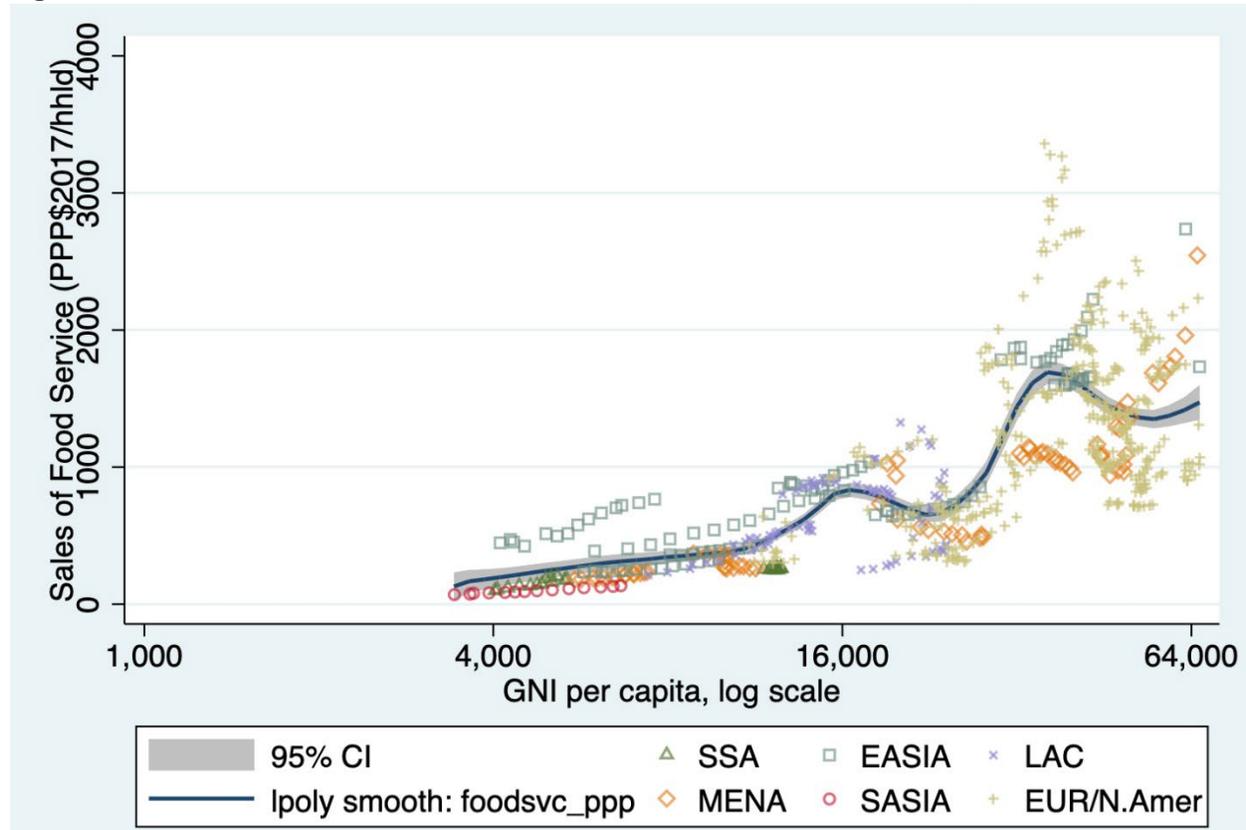

Note:  Data are from Euromonitor for 53 countries from 2006-2019. Both sales and GNI are in constant 2017 US dollars at purchasing power parity prices. The solid line is a local polynomial regression with its 95% confidence interval for the global mean at each income, computed using Stata's lpolyci with default bandwidth.

### 3.4 Demand for packaged foods and soft drinks by income level in Nepal and Bangladesh

The data presented so far aims to describe nutrition and dietary transition using the widest possible range of observations over time and across countries. For economists concerned with international development, looking within low-income countries to compare food choices among poorer to richer households provides a particularly useful kind of evidence. This section presents data from household surveys in Bangladesh and Nepal to describe today's earliest stages of dietary transition in demand for prepared and packaged foods, and for sweetened beverages. We show that the income elasticity of demand for these types of products is high even among the poor, and that budget shares devoted to these products is rising rapidly over time at each income level.

The data for Bangladesh were derived from the Bangladesh Aquaculture-Horticulture for Nutrition Research (BAHNR) study conducted between 2016-2017 in three bi-annual waves of 3060 households. Figure 17 presents non-parametric Engel functions using the first and third rounds of this survey, separated by one year. The top panel focuses on prepared and packaged foods (PPFs), while the bottom panel focuses on sugar-sweetened beverages (SSBs). While the budget shares of total household expenditures for both categories of products are small in absolute terms, it's also clear that those budget shares are increasing significantly, and at a pace that's observable over the course of a single year. The positively sloped segments of these Engel functions indicate that these are luxury goods for approximately the bottom quartile of the income distribution. Indeed, the implied income elasticities of demand remain close to unity throughout the distribution, suggesting that demand will increase in direct proportion to income growth.

Household surveys conducted in Nepal in 2013 and 2016 show similar demand characteristics for PPFs and SSBs (also with implied income elasticities near to or greater than unity), and similar increases over time in the budget shares devoted to those products (Figure 18). The survey rounds in Nepal are separated by three years. Compared with Bangladesh, the budget shares are greater for both types of goods, and while the increases over time in Bangladesh were concentrated at the low end of the income distribution, the increased budget shares for PPFs and SSBs observed in Nepal are concentrated in the middle of the distribution.

**Figure 17. Expenditure shares for packaged foods and soft drinks at each level of total household expenditure in Bangladesh, 2016 and 2017**

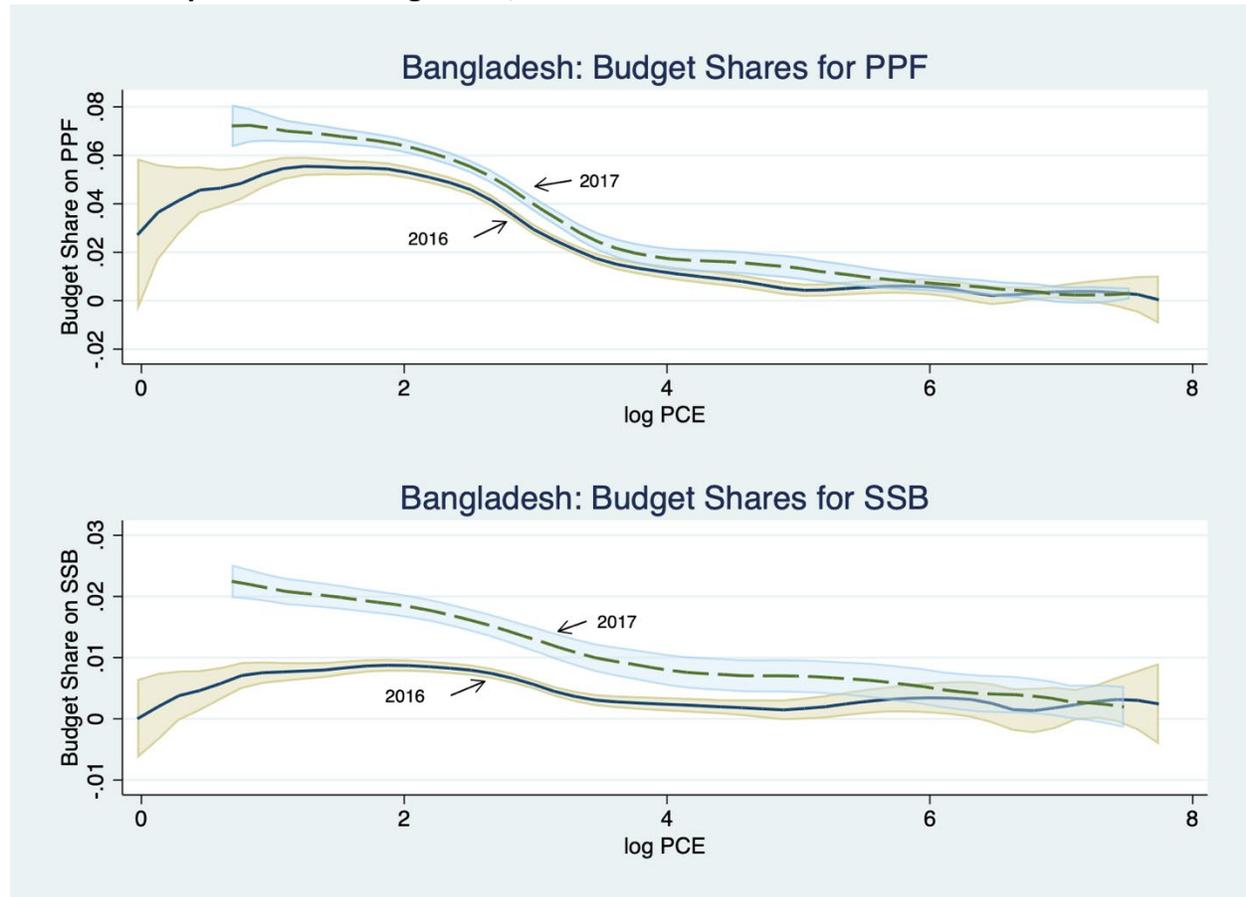

Notes: Household survey data from the Bangladesh Aquaculture-Horticulture for Nutrition Research (BAHNR) study conducted between 2016-2017 in three bi-annual waves of 3060 households. Data shown are for all packaged and processed foods (PPF) and all sugar-sweetened beverages (SSB). The solid and dashed lines are local polynomial regressions with its 95% confidence interval for the global mean at each income, computed using Stata's lpolyci with default bandwidth.

**Figure 18. Expenditure shares for packaged foods and soft drinks at each level of total household expenditure in Nepal, 2013 and 2016**

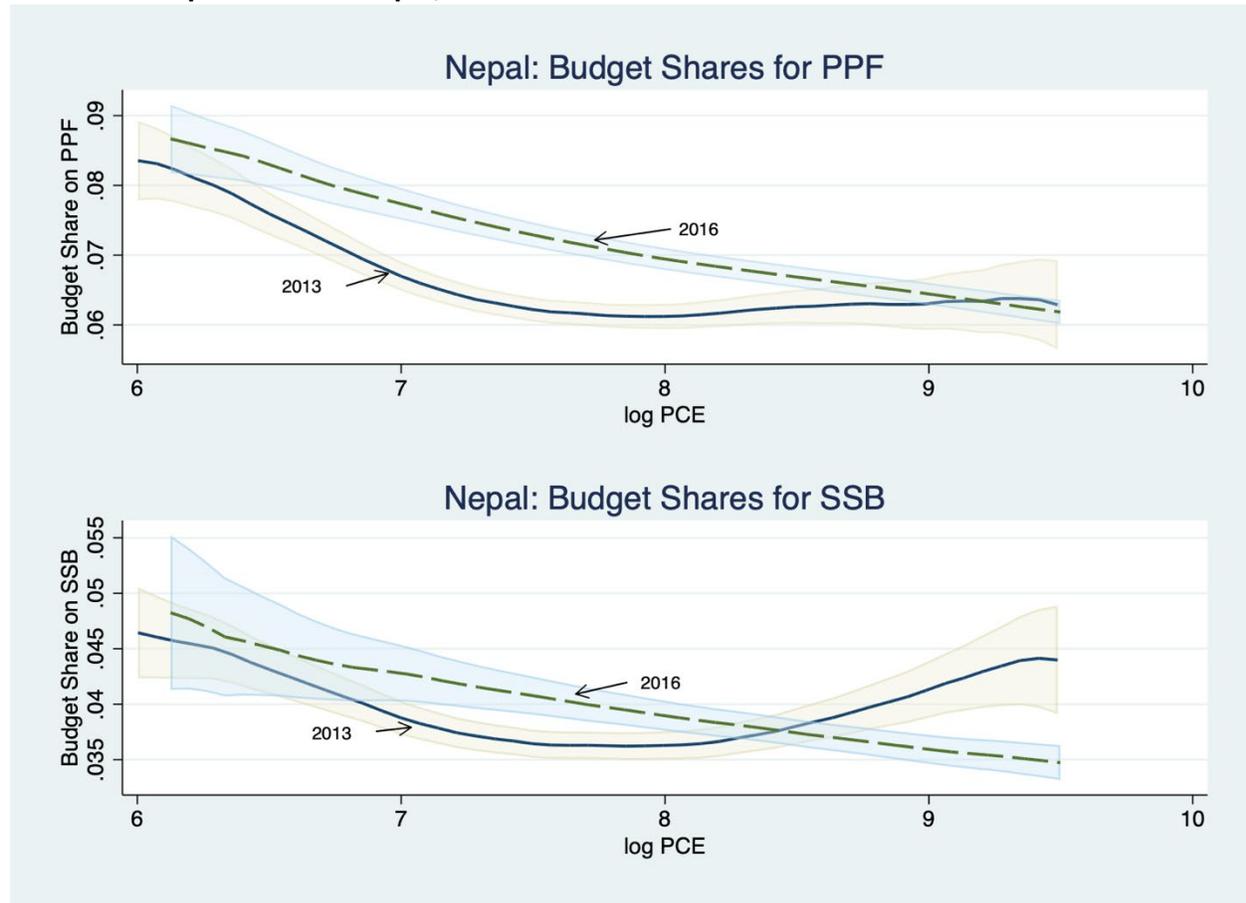

Note: Household survey data from the Policy and Science for Health Agriculture and Nutrition (PoSHAN) Community Studies survey of 40,000 person-visits conducted during four annual, nationally-representative surveys (2013-2016) in Nepal. Data shown are for all packaged and processed foods (PPF) and all sugar-sweetened beverages (SSB). The solid and dashed lines are local polynomial regressions with its 95% confidence interval for the global mean at each income, computed using Stata's lpolyci with default bandwidth.

A crucial question for diet quality is whether increased intake of packaged foods and sweetened beverages displaces starchy staples, or whether they displace more nutrient-dense foods. While a thorough analysis of that question would require estimation of complete demand systems, these categories of food expenditures are broad enough that we can derive suggestive evidence by observing changes in the composition of household food budgets at increasing levels of expenditures on PPFs. Figure 19 illustrates changes in the shares of household food expenditures on cereals and micronutrient-rich foods in Nepal at each level of expenditure on PPFs, while controlling for total household expenditure.

**Figure 19. Share of food budget on cereals and nutrient-rich foods at each level of packaged food expenditure, controlling for total household expenditure in Nepal, 2013 and 2016**

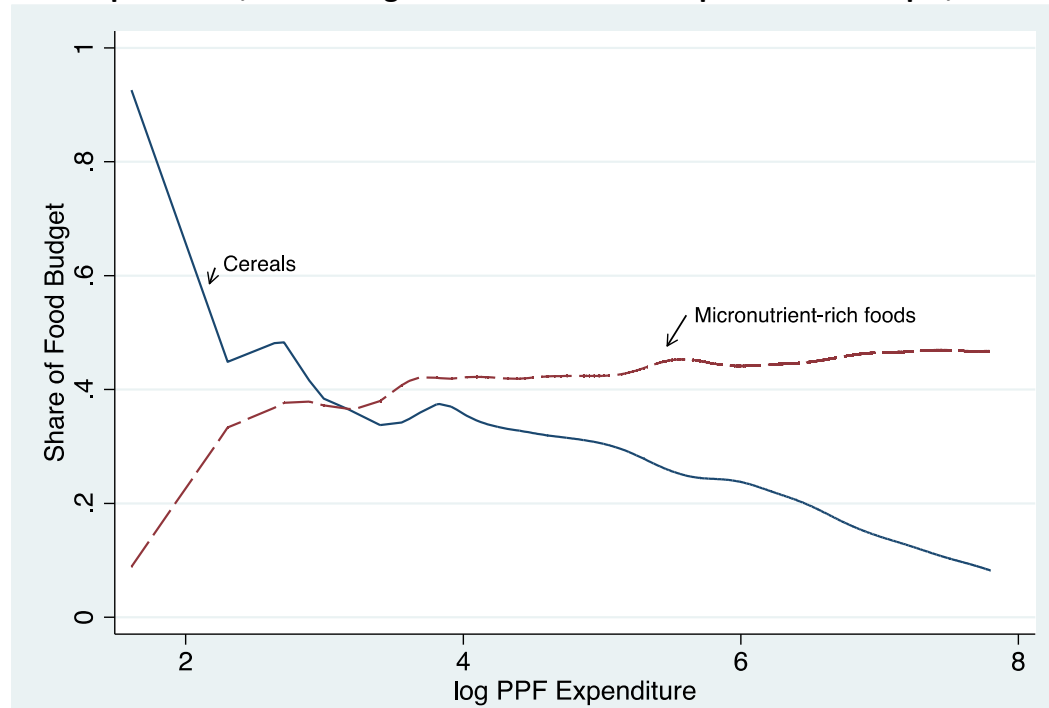

Note: Semi-parametric regression results (using Stata's semipar command), controlling for survey round and region of Nepal, using household survey data from the Policy and Science for Health Agriculture and Nutrition (PoSHAN) Community Studies survey of 40,000 person-visits conducted during four annual, nationally-representative surveys (2013-2016) in Nepal. Data indicate shares of household food budgets allocated to cereals and to micronutrient-rich foods (dairy, animal-sourced foods, and dark green leafy vegetables).

Results shown in Figure 19 use semi-parametric regressions, controlling linearly for total household expenditures per capita, as well as fixed effects for district and survey round. At higher levels of expenditure on PPFs along the horizontal axis, the share of expenditures on cereals is lower and expenditure share on micronutrient-rich foods is higher, primarily at the very lowest levels of expenditure on PPFs. Taken together, results in Figures 18 and 19 suggest that Nepali households in these surveys substitute the "empty" calories in snacks and sweetened beverages for calories in cereal grains, rather than micronutrient rich foods.

## 4. Policies to address the multiple burdens of malnutrition

Programs and interventions can improve nutrition in many ways, starting with "nutrition-specific" actions that target particular foods, nutrients or other compounds, and then broader "nutrient-sensitive" actions that address the underlying determinants of malnutrition such as agricultural production and trade, social safety nets, early child development, maternal education and schooling (Ruel and Alderman, 2013). Nutrition-specific policies and programs are primarily undertaken in the health system, with a set of recommended interventions documented for example in the World Health Organization's Global Database on the Implementation of Nutrition Actions (WHO, 2021), whereas nutrition-sensitive actions are 'multisectoral' in the sense that they involve all parts of the economy and society that affect agriculture and the food system (FAO, 2021). Strategies to estimate the impact of nutrition interventions and policies vary depending on the discipline (Caputo and Just 2022).

Nutrition-specific actions aim to address health behaviors such as exclusive breastfeeding to six months of age, and to address nutritional deficiencies with fortification, supplementation and other actions that address nutritional influences on health and well-being not reflected in consumers' revealed preferences. These interventions can have large effects on public health and economic development, as demonstrated in the large impacts of iodized salt introduced to the U.S. in the 1920s (Adhvaryu et al. 2020) and more recently around the world (Alemu 2022). In contrast, nutrient-sensitive actions typically aim to shift supply and demand for existing foods and have the greatest impact on health when they raise real income for those at greatest risk of malnutrition, lower the real cost of more nutritious foods, and support women as addressed in detail by Quisumbing and Doss (2022).

Nutrition-sensitive actions include education, especially girls' schooling and community outreach to improve conditions for maternal health and child development. Ruel and Alderman (2013) examined the association between parental schooling and child nutrition outcomes using 19 datasets from the Demographic and Health Surveys. They found significant reductions in the risk of child stunting among mothers with at least some primary schooling, with greater reductions coming with higher levels of school attainment. They cite five potential pathways through which maternal schooling may improve child nutrition in addition to the direct effect of greater parental education on household income. These other pathways include: 1) direct transmission of health and nutrition knowledge, 2) numeracy and literacy as vehicles for learning about nutrition, 3) increased receptivity to modern medicine through exposure to new environments, 4) enhancement of women's roles in decision making, and 5) formation of women's social networks. In a more focused study of households in rural central Java, Webb and Block (2004) distinguished female schooling from nutrition knowledge, finding that the positive effect of formal schooling on child nutrition was specific to longer-term outcomes (proxied by HAZ), while the benefits of mothers' nutrition knowledge was specific to shorter-term outcomes (proxied by WHZ).

**4.1 Nutrition-sensitive agricultural interventions**
Agriculture supplies foods and ingredients needed to alleviate malnutrition and also provides employment and livelihoods for many malnourished people, so many nutritional interventions focus on agricultural production. Ruel, Quisumbing, and Balagamwala (2018) provide a comprehensive review of nutrition-sensitive agricultural program evaluations across categories including biofortification, homestead food production, livestock-oriented programs, and nutrition-sensitive value chains based on a total of 45 studies. They find that work on biofortification concentrated on vitamin A, with a focus on children's intake. Biofortification applies agricultural technology to elevate levels of micronutrient levels in food crops. By the end of 2020, targeted breeding had produced 393 fortified crop varieties which had been released or were in testing in 63 countries (Virk et al., 2021). Studies from Mozambique and Uganda summarized by Ruel et al. (2018) found positive effects on children's vitamin A intake and serum retinol levels resulting from training programs on the benefits of consuming orange-fleshed sweet potatoes and distribution of vines for their production. Finkelstein et al. (2019) provide a meta-analysis of studies on iron biofortification with a focus on health outcomes (moving beyond previous studies' assessment of iron biofortification on biomarkers alone). Based on RCT studies of iron-biofortified rice in the Philippines, pearl millet in India, and beans in Rwanda, they report no significant effects on anemia or iron deficiency outcomes among the study populations. Studies did, however, indicate improvements in cognitive functions resulting from consuming iron-biofortified foods.

For biofortification to play its role in supporting daily intake of micronutrients, milestones cited by Ruel and Alderman (2013) include meeting minimum concentrations for each micronutrient through improved breeding, bioavailability of those micronutrients, and farmer adoption rates and adequate consumption by target populations. Van Der Straeten et al. (2020) argue that staple crop biofortification through gene stacking, combining conventional breeding and metabolic engineering, could greatly accelerate progress in reducing micronutrient malnutrition.

Ruel et al. (2018) also review eight evaluations of homestead food production interventions in Burkina Faso, Zambia, Nepal, India, and Bangladesh. These programs often targeted women with provision of inputs for home gardens, training, and in some instances nutrition education. In Burkina Faso, for example, a two-year intervention improved hemoglobin and reduced anemia in children 3-6 months of age at baseline, and reduced wasting and diarrhea, but not stunting, among children 3-13 months of age at baseline. In addition, this study found increased intake of nutritious foods and greater dietary diversity among women participants. A second phase of the study in Burkina Faso further demonstrated even greater reductions in child anemia when these initial interventions were combined with a water, sanitation, and hygiene intervention.

Livestock-oriented programs have tended to prioritize poverty reduction over specific nutrition objectives. Ruel et al. (2018) recount a two-year longitudinal evaluation of a livestock-oriented community development program in Nepal in which rural women received pairs of goats and training. Benefits were found to include improvements in child height-for-age scores. A follow-up study in Nepal by Darrouzet-Nardi et al. (2016) found improved dietary diversity among participant households in those regions better suited to livestock, but not in regions more suited to crop cultivation.

Nutrition-sensitive interventions in value chains may be an important complement to improved farm production. For example, Ruel et al. (2018) cite the example of a dairy value chain intervention conducted among pastoralists in northern Senegal. The intervention tested the efficacy of distributing micro-nutrient fortified yogurt produced using milk supplied by local dairy farmers. An evaluation found a statistically significant increase in hemoglobin among boys (only) and suggestive evidence of a 20-percentage point reduction in the prevalence of anemia among children in the treatment group.

The observational studies reviewed by Ruel et al. (2018) focused primarily around two lines of enquiry -- the relationship between crop production diversity and nutrition outcomes, and the impact on health and nutrition of livestock keeping and sanitation. For populations with limited market access, the degree of nutritional non-separability between a household's food consumption and their own farm production can be summarized by testing for links between their dietary diversity (typically measured as the number of distinct food groups, which is closely linked to nutrient adequacy) and their production diversity (typically measured as the number of different species cultivated). Their general takeaway from the studies that address this relationship was a positive association between crop production diversity and dietary diversity with the important caveat that these associations are stronger in locations more remote from markets or with imperfect market infrastructures. Studies in low-income settings often find a small positive correlation, but effects are frequently limited to specific subgroups (Sibhatu and Qaim 2018), and studies that link consumption to production of specific foods often find effects only for specific foods such as eggs or dairy that are less easily traded (Mulmi et al. 2017). Linkages between crop diversity and nutritional status were generally found to be weaker.

The livestock studies reviewed by Ruel et al. (2018), addressed both production-consumption linkages (for dairy products in particular) and the potential health and sanitation implications of exposure to livestock waste and diseases. Studies from Ethiopia, Uganda, Tanzania, and Nepal found positive associations between dairy production and milk consumption, with attendant improvements in child anthropometric outcomes. Yet, other studies noted the need to weigh these benefits against the health risks of exposure to livestock. Studies in Bangladesh, Ethiopia, and Vietnam found negative associations between the presence of animal feces in household compounds and child HAZ.

Agricultural interventions can also promote improved nutrition by lowering food prices through cost-reducing innovation, supported by public-sector R&D, infrastructure and institutions. Agricultural interventions for staple foods have proven efficacy in raising incomes in resource-poor regions (Gollin, Hansen and Wingender 2021), with important consequence for child health and survival (von der Goltz et al. 2020). Gollin et al. (2021) find that high-yielding varieties developed in the Green Revolution increased yields by 44% between 1965 and 2010, increasing income and reducing population growth. Von der Goltz et al. (2020) add that diffusion of modern varieties reduced infant mortality globally by 2.4-5.3 percentage points. Anderson and Birner (2020) caution that these benefits of the Green Revolution were primarily in staple grains and starchy food crops, largely to the exclusion of fruits and vegetables.

## 4.2 Safety-net transfers and nutrition assistance
Many governments and aid groups offer some sort of safety net assistance to low-income people, often through cash or vouchers and in-kind transfers designed to smooth food consumption at times and places of greatest need. Social safety nets have also proven effective tools to promote maternal and child nutrition. Conditional cash transfer programs provide targeted households (often mothers specifically) cash payments in return for specified practices or behavioral changes aimed at enhancing children's health, education, and nutrition (de Groot et al. 2017). Most of the experience and evidence relating to conditional cash transfer programs in developing regions comes from countries in Latin America. Experience in Sub-Saharan Africa has been sporadic, as Onwuchekwa, Verdonck, and Marchal's (2021) systematic review found only three studies reporting nutritional impacts of CCTs in the region, with only one showing a positive effect. Afridi (2010) reports substantial increases in daily nutrient intake among children participating in a mandated school meal program in India.

Latin American examples of national-scale nutrition oriented conditional cash transfer programs include Mexico's *Prospera-Oportunidades-Progresa*, Brazil's *Bolsa Familia*, and Colombia's *Más Familias en Acción*. Conditions in each of these programs included child vaccination and growth monitoring, prenatal care (in some cases including health and nutrition class attendance), and thresholds for children's school attendance, all targeted to households living in poverty or extreme poverty. The most intensely evaluated of these programs is Mexico's, which has generally been found to have had positive impacts on nutrition outcomes among the target population (Gertler 2004; Rivera et al. 2004; Barber and Gertler 2008; Leroy et al. 2008; Huerta 2006). This program ran from 1997 to 2019, with the goal of improving children's access to schooling and households' access to primary health and nutrition services by providing cash incentives to female heads of household (Segura-Pérez, Grajeda, and Pérez-Escamilla 2016).

Evaluations of *Progresa's* impact on child nutrition have found that in utero and early childhood exposure to the program was associated with reduced morbidity, 1 cm increased height in children 12-36 months of age, reduced prevalence of anemia, reduced prevalence of stunting, and increased height-for-age Z-scores as compared with children in non-participating households. Evaluations of *Progresa* have also found increases in secondary school enrollment rates, increased participation in preventive care and prenatal health visits, and increased household food expenditure (Segura-Pérez, Grajeda, and Pérez-Escamilla 2016). Hoddinott and Skoufias (2004) found that *Progresa* increased caloric intake among participating households, particularly from fruit/vegetable sources and from animal sourced foods. A more recent experimental evaluation of *Progresa's* impact on nutrition produced more detailed results (Kronebusch and Damon 2019) regarding specific micro- and macro-nutrients. Kronebusch and Damon find both positive and negative nutrition effects. Among the former, they find especially positive effects of *Progresa* on micronutrient intake, with increases in vitamin A, vitamin C, and iron (and other essential minerals). However, they also find that treated households also increased their consumption of processed carbohydrates, sugar, and saturated fat.

A review of 17 analyses of Brazil's *Bolsa Familia*'s impact found a lack of consensus regarding program effects on food security, nutrition status, food acquisition, and quality (Neves et al. 2020). Several studies reviewed found evidence of increased food access among households participating in *Bolsa Familia*. Other studies differed in their findings regarding whether, or to what extent, children in participating households showed improved nutritional outcomes. Neves et al., summarize that a national survey found improvements in weight-for-age among children in participating households, while other studies found no or negative effects on height-for-age. A review of *Más Familias en Acción* in Colombia found increases in HAZ scores and decreased probability of stunting among children 0-24 months of age (Attanasio et al. 2005).

The United States has one of the world's oldest and largest nationwide food safety net systems, with a specialized program for Women, Infants and Children (WIC) that starts in pregnancy, and a general Supplemental Nutrition Assistance Program (SNAP) designed to serve everyone with sufficiently low levels of income and wealth. The economics literature on the health impacts of these and other programs is surveyed in Bitler and Seifoddini (2019). As shown by Hoynes, Schanzenbach and Almond (2016) and more recently by Bailey et al. (2020), the rollout of SNAP in the 1960s and early 1970s is linked to significant improvements in long-run educational and economic outcomes, especially for those who were young children when the program was introduced. WIC benefits allow users to buy target quantities of specific healthy foods, while SNAP funds can be used to buy almost any food or beverage except alcohol and hot prepared items. The quantities offered through WIC are designed to meet all of an infant's needs, while the benefit level offered in SNAP is designed to supplement the household's own spending on groceries. Multiple surveys have confirmed that the program works as intended, with recipients spending more than the benefit level on SNAP-eligible foods (Hastings and Shapiro, 2018).

As long as SNAP recipients spend some of their own money in addition to program benefits on eligible foods, they would have little or no incentive to traffic their benefits and convert benefits into cash. Most studies find that SNAP is inframarginal in this sense, working as intended to provide supplemental assistance rather than the entirely of what a consumer would spend on food. But SNAP benefits cannot be spent on hot prepared food, so its primary effect on diet composition may be to shift consumption towards groceries and packaged foods rather than food away from home. Hasting and Shapiro (2018) estimate that the marginal propensity to spend SNAP benefits on groceries and packaged foods is 0.5-0.6, significantly higher than the value of 0.1 they estimate for otherwise similar U.S. households. Other studies using different approaches also find that beneficiary households obtain more of their food from SNAP-eligible cold items for consumption at home, than comparable non-SNAP households whose consume more hot prepared items and food away from home (Ismail et al. 2020).

Hastings, Kessler, and Shapiro (2021) examine the effect of SNAP with respect to the share of calories devoted to various food groups, as well as the effects relative to the FDA's nutrient density score and the USDA's Healthy Eating Index. Their finding is consistent with the benefits being inframarginal and non-distorting among groceries, although they observe slight reductions in the share of kilocalories coming from fruits and vegetables, and a slight increase in the share coming from total fat, which might be due to SNAP being disbursed only once per month leading to purchase of less perishable items. They further find small reductions in the healthfulness of spending by SNAP households relative to indicators of nutrient density and the Healthy Eating Index. Hudak and Racine (2019) summarize the mixed evidence regarding the impact of SNAP on child weight with the conclusion that participation helped boys to maintain normal body weight, but SNAP may have contributed to obesity and overweight among girls.

Social safety net programs include food provided at childcare and educational facilities, as well as the health care system, food pantries and other distribution programs. School meals are typically designed primarily to achieve child development and educational objectives, but may have significant nutritional impacts as well. Example studies of large, national preschool programs include work on India's Anganwadi centers (Mittal and Meenakshi, 2019) and the U.S. Head Start program (Carneiro and Ginja 2014), both of which are found to have significant impacts on nutritional outcomes. Positive nutritional effects of meals provided for older children have also been found in a variety of settings such as Ghana's national school meals program (Gelli et al, 2019).

Some countries provide food with financing through the health care system under initiatives known as food by prescription or medically tailored meals, and many communities provide donated food items through a variety of distribution channels. The nutritional composition of these transfers can potentially have a large-scale impact, whether they serve targeted groups at congregate sites such as soup kitchens, home delivery services or food banks that distribute donated and purchased groceries through food pantries (Gundersen et al. 2021). These transfer programs using donated food can account for a large share of total food consumption for some beneficiaries, and pose complex problems of how to meet needs effectively in the absence of price signals (Prendergast 2017; Byrne and Just 2021).

### 4.3 Taxes and subsidies
Almost all governments use many tax and subsidy instruments to alter relative prices and target government revenue towards consumption and production of specific things. Longstanding taxes include a variety of excise or surtaxes on items to be discouraged, and subsidies often take the form of targeted exemptions from general sales tax and value-added tax paid when items are sold, or exemptions from taxation of producers' income, capital gains, real estate and use of specific inputs in food production. Tax and subsidy interventions are introduced to meet diverse and often conflicting goals, creating opportunities for reform towards coherence in pursuit of new objectives such as improved nutrition (OECD 2021).

Taxes and subsidies affect nutrition in surprising ways, in part because of differences in market structure that affect retail food prices as opposed to agricultural commodities. For example, in the U.S. since 1985 consumer prices of fruits and vegetables have been rising while prices of sugar-sweetened beverages have been falling (Christian and Rashad 2009), reflecting determinants of retail prices that are independent of major commodity price interventions such as trade agreements that lower U.S. fruit and vegetable prices through more imports from Mexico (Johnson 2017), or import restrictions that raise U.S. prices of raw and refined sugar (Sinclair and Countryman 2019).

In response to the obesity epidemic, public health authorities often seek specific taxes on harmful foods to reduce their consumption and raise revenue for targeted initiatives. Despite an early finding by Cash, Sunding, and Zilberman (2005) that subsidies on fruits and vegetables are both a relatively cost-effective intervention with regard to the cost of a statistical life saved and distributionally progressive, taxes on unhealthy foods and ingredients have received greater attention and application. Particular attention has focused on taxation of sugar-sweetened beverages (SSBs). As of 2021, over 45 countries (or localities) have introduced taxes on SSBs, with additional examples of countries imposing taxes on other food categories such as ice cream, confectionaries, and foods high in saturated fats (Popkin and Ng 2021).

Taxes on SSBs and other potentially harmful foods can be *ad valorem* as a percent of the sales price, or a specific value per unit sold. For products with harmful ingredients, taxes may be levied per unit of that ingredient or tiered to encourage product reformulation and marketing changes (Lacy-Nichols, Scrinis and Carey 2020). To address substitution among unhealthy ingredients, taxes could potentially be based on jointly defined thresholds for example on saturated fat, sugar, and sodium (Caro et al. 2017), or based on overall nutrient profile scores designed to signal a food's association with long-term health (Deschasaux et al. 2020).

The impact of taxes or subsidies on consumption depends on how it influences product formulation and marketing efforts as well as the final retail price, which in turn interacts with consumers' response to price changes or reformulation and differences in product marketing, including public health messaging that accompanies rollout of a new tax (Claudy et al. 2021). In cases where there is no reformulation, key questions include the degree to which the tax raises consumer prices in different retail settings (Cawley et al. 2021).

Systematic reviews of taxes on unhealthy foods, and SSBs in particular, generally find that such taxes work as intended to reduce consumption of the taxed items (WHO, 2015). Even without reformulation and marketing changes, whatever tax is passed through to a price increase is expected to reduce consumption by the same proportion, as SSB demand is estimated to be roughly unit-elastic (Teng et al. 2019). For the fraction of consumers who respond with reduced intake, the resulting health gains offset their loss of enjoyment and consumer surplus, but some consumers continue high levels of consumption and pay a substantial cost. These distributional concerns lead many tax laws to earmark the revenue for health and other services (Krieger et al. 2021).

The first nationwide tax on SSBs designed specifically in response to rising obesity rates was introduced by Mexico in 2014, at 1 peso per liter or approximately a 10 percent price increase (Backholer et al. 2016). The Mexican government took this step, despite great opposition from food and beverage companies, due to having the world's highest level of SSB consumption averaging 163 liters per capita per year that accounted for 70 percent of the country's added sugar consumption, with prevalence of overweight or obesity at 30 percent of children and 70 percent of adults (PAHO, 2015).

Multiple studies have confirmed the success of Mexico's SSB tax in suppressing consumption. Nearly the entire amount of the tax was passed through to retail prices. After the first year of implementation, an initial review found a 6 percent reduction in purchases of SSBs, with a larger reduction (17 percent) among consumers in the lowest income quintile (PAHO, 2015). Purchases of non-taxed beverages, including milk and bottled water, increased 7 percent in the first year of the SSB tax. This tax also resulted in significant increases in government revenue ($533 billion between 2014-17), some portion of which was allocated to the installation of public fountains for drinking water in schools and public places. A review conducted after the second year of SSB taxation found that purchases of taxed beverage fell by 5.5 percent in the first year, and by 9.7 percent in the second year, with 2-3 times greater reductions in SSB purchases in the lowest as opposed to highest socio-economic group (Colchero et al. 2017). After the third year, another study found that the probability of respondents being a medium- or high-level consumer of SSBs fell by nearly 13 percentage points, while the probability of consuming none or a low level of SSBs rose by nearly 5 percentage points (Sánchez-Romero et al. 2020).

A key aspect of Mexico's SSB tax is the signaling that accompanied public debate and rollout of the new tax. Álvarez-Sánchez et al. (2018) examined the impact of consumers' awareness of the SSB tax on purchasing, and found that consumption declined nearly 16 percent more among those consumers who were aware of the tax than among those who experienced the higher price but did not report knowing about the tax. This informational spillover is the counterpart to the advertising effects of subsidy programs, for example when SNAP recipients in the U.S. who received healthy incentive pilot rebates on some fruit and vegetable purchases increased consumption of eligible items, and also purchased more of other fruit and vegetable items for which no discount was offered (Wilde et al. 2016).

The distributional effects of SSB taxes remain controversial, with mixed evidence from different settings. The Mexican finding that the lowest socio-economic groups were most willing and able to substitute away from SSBs and therefore gained more of the health benefits is substantiated by a study in Australia (Lal et al. 2017), but contradicted by a study in Indonesia which found substantially greater reductions in obesity and diet-related NCDs among higher income quintiles (Bourke and Veerman, 2018). A study in South Africa found the greatest health benefits in the middle income quintiles (Saxena et al. 2019). All three of these studies found substantial savings in government health care expenditures from the reduced consumption, which could be used along with tax revenue from continued consumption to help compensate the most affected communities. In the U.S. context, both the internality and externality benefits among those who reduce consumption and spending of tax revenue could be made highly progressive (Allcott, Lockwood, and Taubinsky 2019).

Another economically important aspect of SSB taxes and other health interventions is the age profile of effects, because impacts earlier in life typically lead to larger total impacts per person. A study in the UK found that SSB taxes were more effective at reducing consumption in younger consumers (Dubois, Griffith, and O'Connell, 2020), although a study in Mauritius found no significant difference by age (Crawley, Daly, and Thornton 2021). The Mauritius study is among the few studies of SSB taxes with a strong emphasis on causal identification, and also interesting subgroup analysis finding a significant reduction in consumption for boys but not girls (Crawley, Daly, and Thornton 2021).

Cost-effectiveness studies of the Mexican SSB tax data find it to yield substantial savings to the government, reducing health care costs by over 90 million U.S. dollars over 10 years (Basto-Abreu et al. 2018). That study and others in the health literature focus on health outcomes, for example finding that by 2024, the SSB tax of 1 peso per ounce would prevent over 61,000 cases of diabetes, nearly 700 cases of cancer, and approximately 11,000 cases of stroke and heart disease. Relatively few students attempt similar estimation of economic welfare effects, taking account of consumer and producer costs given their substitution possibilities as done for the U.S. by Lee and Giannakas (2020).

Research on price instruments to improve nutrition typically focuses on taxes to reduce harms from products like SSBs in upper-middle and higher-income countries where they are already widely consumed. Taxes on unhealthy foods in lower-income countries receive less attention, perhaps reflecting the political economy of SSB taxes as shown in a review of seven countries in Sub-Saharan Africa where industry-led opposition frequently blocked adoption and implementation (Thow et al. 2021). In lower-income settings the food industry accounts for a larger fraction of GDP, employment and consumer expenditure than in higher-income countries, and diet-related diseases often go untreated. This results in less political pressure for prevention. Conditions like in Mexico have arisen also in South Africa, where the Health Promotion Levy introduced in 2018 could help limit SSB consumption (Stacey et al. 2019) and achieve significant health improvements with relatively little loss of consumer welfare for the poorest (Saxena 2019).

A final category of price instruments includes subsidies for healthy items. These are typically limited to grocery sales tax exemptions (Zheng et al. 2021) or rebates and discounts on certain items purchased with nutrition assistance (Wilde 2016). Broad subsidies on all fruits and vegetables like those considered in the early study of Cash, Sunding, and Zilberman (2005) are not typically implemented, perhaps due in part to the great diversity of items and unclear boundaries on what would be included along the spectrum of more to less healthy items. Even within each species of fruit or vegetable there is a high degree of store-specific price dispersion and variation over time relative to wholesale costs, suggesting low impact of subsidies on retail pricing (Lan and Dobson 2017). Efforts to increase use of healthy items therefore consist primarily of transfer programs to consumers discussed in the previous section, or systemic interventions to lower costs by improving production and distribution of healthy foods, as well as regulatory approaches described below.

**4.4 Regulation and labelling**
The fact that consumers cannot see, taste or smell the nutritional value of foods, especially when they are newly introduced and not yet subject to long periods of experimentation and selection into culinary traditions, ensures that information about food composition and its impact on health can potentially be used to remedy market failures and improve outcomes. Regulation of food production and marketing can take a variety of forms, ranging from outright bans on specific processes, products or ingredients to rules and standards regarding food manufacturing and inspection, product labeling and advertising.

Among the most remarkable examples of public action using regulatory approaches to improve nutrition is the story of industrial trans fat. This ingredient first entered the U.S. food supply to stabilize margarine in 1910, and gradually rose in use to over 30 lbs/year per person as a widely used ingredient and retail product sold for home use (Crisco). Trans fats were then found to be harmful and required to be disclosed on product packaging after 2006, and ultimately banned entirely from sales within the U.S. after 2018 (Amico et al. 2021). By late 2020, a total of 58 countries had also introduced laws limiting or eliminating this ingredient, which is thought to be responsible for approximately 500,000 deaths annually from coronary heart disease (WHO 2020).

Trans fats could enter and exit from the U.S. relatively quickly because it has close substitutes. Regulation of other ingredients that are now known to be harmful at high levels such as sodium, sugar and refined carbohydrates (as opposed to whole grains) are almost entirely now subject only to disclosure rules, but could potentially be regulated in terms of ratios or relative quantities. Regulatory approaches to food composition mirror rules now used for older products known to be dangerous when consumed to excess, particularly alcohol, and are in many ways an extension of food safety rules pioneered in the U.S. and other high-income countries that are now spreading around the world (Hoffman, Moser and Saak 2019).

Nutritional disclosure and labeling in the U.S. originated from recommendations of the 1969 White House Conference on Food, Nutrition, and Health (Institute of Medicine, 2010). When formally adopted in 1973, the regulations stipulated that nutrition labeling include numbers of calories, grams of protein, carbohydrates, and fat along with percent of recommended daily allowances of various micronutrients and protein. The law left to manufacturers' discretion whether to report the level of sodium, saturated fat, and polyunsaturated fatty acids. Later the U.S. Congress significantly expanded and updated these labeling requirements with the Nutrition Labeling and Education Act of 1990. This law required nutrition labeling on most packaged foods and provided guidelines for how labels could describe nutrient content relative to daily diet, for how serving sizes could be expressed, and extended voluntary labeling for raw fruits, vegetables, and fish. Similar labeling regulations in the European Union followed only in 2016. Broader international efforts to improve food labeling could be global in nature, working through the Codex Alimentarius initiative of the United Nations (Thow et al. 2020).

Even if labels clearly communicate the nutritional composition and health effects of foods, they may have limited effects on food choice, as shown by a meta-analysis of the effects of food labeling on consumer diets and industry practices: In 60 studies across 11 countries, labeling was successful in reducing consumption of total energy and total fat, and increasing consumption of vegetables, but had no significant effects on consumption of sodium, total carbohydrates, protein, saturated fats, fruits, of whole grains (Shangguan et al. 2019). Regarding industry practices, they found that food labeling influenced formulations for trans fat and sodium, but not for total energy, saturated fat, or dietary fiber.

Over time, the focus of disclosure requirements has shifted from ingredient lists and nutrient fact panels listing specific items to overall food profile scores summarized in a front-of-package label. Chile was the first to adopt a nationwide mandate in 2012, requiring a kind of traffic light signal to indicate levels of unhealthy ingredients. By 2018, four other countries in South

America had followed suit. These early labels provide only negative warnings against potentially harmful foods, while other countries are introducing traffic lights, multiple stars or numerical scores to distinguish healthier foods from dangerous ones (Crosetto et al. 2020).

Beyond labeling, many countries regulate what can be provided or sold around specific institutions such as schools. For example, Bergallo et al. (2018) survey rules on SSB sales near schools in Latin America, finding a near total ban in Mexico, a ban on drinks with added sugar greater than 2.5g/100ml in Peru, and a ban on drinks with greater than 7.5g/ml in Ecuador and Uruguay. Restrictions on private sale of other items that would compete with healthy foods are an important part of school food policy in the U.S., pioneered in Massachusetts in 2012 and later extended nationally (Jahn et al. 2018).

Finally, an important and longstanding form of regulation concerns advertising, especially when targeted to children. The scope and nature of these interventions varies widely, from complete bans on advertising items such as SSBs in any media, to prohibitions in particular outlets such as TV and movie theaters (in the Mexican case, excluding sporting events and soap operas). In countries such as France, regulations require SSB advertisements to include messages to promote healthy lifestyles or to warn about health risks and obesity, while countries such as the UK use a code of conduct found to have a large impact on advertising intensity associated with a significant shift in food expenditure from potentially harmful foods to fruits and vegetables (Silva, Higgins and Hussein 2015).

### 4.5 Goldilocks policy: towards adequacy without excess

A central challenge in the economics of nutrition is that many food components have a U-shaped relationship to health and longevity. Modern societies have a long history of scarcity and want more, but can now overshoot so that goods become bads. For example, refined flour offers extended shelf-life and other features, with little harm until simple carbohydrates from all sources including sugar and potatoes are consumed in large doses without the moderating effect of fiber and other macronutrients.

To help countries achieve adequacy without excess, the 2015 Global Nutrition Report (IFPRI, 2015) coined the term "double-duty actions" that simultaneously target both stunting and obesity (e.g., the double burden). Subsequent efforts have extended the concept towards healthier diets that are also environmentally sustainable regarding climate change, water use and pollution, biodiversity and other national resources such as the EAT-Lancet Commission report (Willett et al. 2019). Analysis of links between climate change and crop yields are well addressed by Ortiz-Bobea (2021), but climate change also affects the nutritional composition of farm output (Smith and Myers 2018). At the efficiency frontier there will be tradeoffs between objectives, but it may be possible to achieve healthier diets that also have low use of natural resources through greater efficiency and equity in global food systems (Hawkes et al. 2020).

Meeting a variety of needs calls for a variety of approaches. In their *Lancet* review, Hawkes et al. (2020) propose 10 candidate actions across 4 sectors of government: health care services, social protection programs, educational institutions, and agriculture or food system governance. Their proposed interventions via health systems include scaled up antenatal care (along with careful monitoring of protein and energy supplements for pregnant women), promotion of breastfeeding, redesigned guidelines for complementary feeding to promote healthy diets, redesigned growth monitoring programs to increase sensitivity to childhood obesity and prevent harm from energy-dense and fortified foods and supplements. Interventions via social protection programs and safety nets focus on cash or food transfers combined with nutrition education and incentives to promote healthy dietary choices. Educational settings provide opportunities to implement school feeding programs that promote healthy diets and regulate marketing of energy-dense snacks and SSBs. Finally, agriculture and food systems offer opportunities to improve peoples' food environment, in the sense of relative prices, availability and marketing. These aim for diversity in food production and consumption, empower women, and highlight urban and peri-urban agriculture, specifically regarding vegetables and fruits, nuts, legumes, and whole grains, along with regulation of breastmilk substitutes and other unhealthy foods that would displace healthier items in an overall balanced diet.

## 5. Themes and conclusions

This chapter describes new information about and understanding of the multiple burdens of malnutrition that have become increasingly important drivers of health outcomes, as well as their implications for how economics research can guide development of agriculture, policy, and food systems around the world. Recent findings reveal rapid change in diet composition, including displacement of traditional foods by new products whose consequences for health are not readily observable, requiring new research and policy intervention to align food choice with each consumer's nutritional wellbeing. Much recent research on dietary transition and food system transformation is conducted in the health sciences, using epidemiological techniques and randomized trials to identify potentially causal links between diet and disease, and biochemical or clinical evidence regarding the mechanisms behind those observations. Malnutrition today is characterized by both deficiencies and excesses in multiple dimensions, including stunted growth in utero and infancy followed by unwanted weight gain and obesity later in life, accompanied by inadequate intake of many micronutrients and other dietary components needed for each life stage including pregnancy and breastfeeding.

A central theme of recent nutritional research has been to identify the early-life origins of adult disease (Barker 1990, Daniels 2016). Much new research on the economics of malnutrition involves interactions between dietary intake and other aspects of human development, using the rapidly growing volume and quality of data and analytical techniques available to study dietary intake and its consequences in both low- and high-income countries. Each kind of data offers distinct strengths and limitations. For agricultural and development economists, the most widely used nationally-representative surveys of individuals and households include National Health Nutrition and Examination Surveys for quantitative information on dietary intake and health outcomes in the United States and a few other countries, Demographic and Health

Surveys that focus on mothers and preschool children in Africa, Asia and Latin America, the Living Standards Measurement Studies and related household expenditure surveys that track consumption spending in relation to other household activities in almost all countries of the world. These surveys and many other recently developed data sources provide researchers with unprecedented power to measure patterns and change in food systems and nutrition, although the high cost and selective focus of existing efforts leaves many blind spots to be addressed by new kinds of data and diverse techniques such as machine learning and Bayesian modeling.

**The role of economics in nutrition research**

Any stylized description of global patterns such as the nutrition transition will oversimplify the actual changes observed at any given place and time. Methodological innovations and new data sources have led to important new insights about the magnitude and significance of quantitative relationships, complemented by qualitative understanding based on familiarity with the people and places that we study. Agricultural economists have a particularly useful toolkit to advance the frontiers of knowledge about food systems and health, based on decades of experience studying the interaction between human behavior, farm production and natural resources. This toolkit includes at least five important ways in which agricultural and other applied economists have helped expand knowledge to drive improvements in nutrition and health:

First, food choices and nutrition can be measured in many ways as summarized in our ABCDEFG mnemonic, resulting in a vast flow of quantitative data suitable for estimation and hypothesis testing. Agricultural economists often have experience using a wide range of time series and cross-sectional or panel data observed at multiple levels of aggregation, building skills and flexibility in database management and data analytics, including innovations in machine learning and computer science as detailed by Baylis (2022). The econometrics toolkit uses economic principles to design and use statistical tests, interpreting each observation as the result of individuals' choices. These techniques are well adapted to analyzing food and nutrition data, identifying how universal aspects of human nature can lead to different outcomes under different circumstances.

Second, diet-disease interactions offer some opportunities for randomized trials, but most data are purely observational. Agricultural economists often have experience combining different kinds of evidence for causal inference, using both experimental and observational data to identify cause-and-effect relationships and infer their generalizability to other settings. Empirical strategies such as instrumental variables, regression discontinuity, matching and synthetic controls or difference-in-difference methods are widely applicable in nutrition research, and used with a variety of falsification tests and placebo regressions or other methods to assess robustness. The causal inference toolkit in econometrics is distinctive in that each observation is interpreted as resulting from individual choices and societal equilibria, as described in this chapter's summary of demand system estimation and revealed preference. This work on food choice complements the causal inference toolkit of health scientists, whose work in biostatistics and nutritional epidemiology is adapted to disease processes that progress physiologically, without economic decision-making.

Third, the economics of nutrition explains food choice as the result of decisions made under uncertainty, in pursuit of diverse goals and a variety of other constraints. The use of constrained optimization to explain individual choices, with equilibria among individuals to explain societal outcomes, provides a set of structural models to predict changes and assess potential outcomes in a unified framework that links health to other individual and social objectives. The use of demand systems and other simultaneous-equation models allows economists to work with multidimensional data, identifying tradeoffs and mechanisms for change. Each model is customized to address a specific set of concerns, but draws on standard economic principles to move towards complete accounting and logical consistency within and between models. Studies of consumer behavior about how changes in prices, incomes, and preferences relate to food choice can be linked to agriculture and food production, distribution and marketing, for a unified view of relations between nature, technology and peoples' choices.

Fourth, economic models link the micro-foundations of individual choices to market structures and macroeconomic phenomena such as inflation and unemployment, providing a wide range of insights into social, commercial and political determinants of health. These insights support accurate measurement of variables such as real versus nominal prices, appropriate functional forms and model specifications for supply and demand. The insights from economic models also reveal how private enterprises respond to incentives in agriculture and food production, reformulation and marketing, and how businesses interact with governments through lobbying and the political economy of food systems.

Finally, economic approaches to food and nutrition are particularly well-suited for the design and evaluation of policies, programs and other interventions. Even for actions that aim purely to improve public health, economic methods can inform how people respond to incentives, can illuminate unintended consequences and can estimate mechanisms for action. As we have outlined in this chapter, many factors interact to determine nutritional status for each individual and every population over time. The framework of dietary transition reveals how diet-disease relationships often have U-shaped functional forms, with a variety of effect modifiers that lead to context-specific impacts. As the literature reviewed in this chapter continues to grow, researchers and decision-makers will be able to draw on an increasingly rich body of knowledge to guide intervention towards sustainable and healthy diets for all.

**Conclusion**

This chapter reviews the economics of nutrition, describing the kinds of malnutrition and diet-related disease that arise during economic development and food system transformation. The central stylized fact we observe is a dietary transition from undernutrition to overconsumption before populations can sustain a balanced diet in terms of food groups, essential nutrients, and other attributes needed for lifelong health. The dietary transition can be observed in average consumption for whole populations over time, and for individuals over the life course. The speed and nature of transition varies greatly, leading to coexistence of multiple burdens of malnutrition between and within countries, households and individuals.

The multiple burdens of malnutrition start with insufficient energy, inadequate nutrients and illness in utero and infancy, leading to stunted growth and impaired child development. Those same individuals often have imbalanced diets leading to excess weight gain later in life, and deficient or excess intake of micronutrients and other compounds that raise risks for a wide variety of metabolic and infectious diseases. The double burden of short stature and excess weight, accompanied by other burdens related to food composition, involve the gradual discovery of physiological and metabolic processes as well as psychological, social and economic responses to rapidly changes in food production, distribution and marketing.

Completing the dietary transition calls for a variety of actions by global and national authorities, institutional and business leaders as well as individual consumers. The need for change is increasingly clear, as poor diet quality is now the largest avoidable cause of death and disability in the world. Other health risks related to economic development provide many historical examples of how change occurs, in domains such as smoking, alcohol abuse, vehicle accidents, water quality, air pollution, house fires, exposure to toxins and occupational injuries. Regarding food and nutrition, the rapidly growing literature reviewed in this chapter provides a useful guide to new data and evidence about the causes of malnutrition, and how policy actions could lead to better outcomes. This research spans multiple disciplines and scholarly outlets, much of which uses the distinctive toolkit of agricultural and applied economists. Future work building on this knowledge could help guide decision-making towards sustained improvements in how food is produced and consumed around the world, leading to significant improvements in diet quality, nutrition and health.